\documentclass{article} 
\usepackage{amsmath} 
\usepackage{jheppub_kim}
\usepackage{longtable}
\usepackage{epsfig}
\usepackage{amsmath}
\usepackage{graphicx}
 \usepackage{graphics}
\usepackage[hang,nooneline,scriptsize]{subfigure}
\usepackage{array}
\usepackage{graphics}
\usepackage{color}
\usepackage{amssymb}
\usepackage{eucal}
\usepackage{enumerate}
\usepackage{graphicx}
\usepackage{epsfig}
\usepackage{amsfonts}
\usepackage{amsmath}
\usepackage{footnote}
\usepackage{multicol}
\usepackage{booktabs}
\usepackage[flushleft]{threeparttable}
\usepackage[font=small,labelfont=bf]{caption}

\title{Quantum-corrected thermodynamics of Schwarzschild AdS black holes in conformal Killing gravity } 
\author{
Saheb Soroushfar \\
Department of Physics, College of Sciences, Yasuj University, 7591874934, Yasouj, Iran \\
soroush@yu.ac.ir
\\[2ex]
Sayyed Mehrab Ramezani \\
Department of Mathematics, College of Sciences, Yasouj University,  7591874934, Yasouj, Iran \\
m.ramezani@yu.ac.ir
\\[2ex]
Hoda Farahani \\
School of Physics, Damghan University, Damghan, 3671641167, Iran \\
h,farahani@umz.ac.ir
\\[2ex]
Saeed Noori Gashti \\
School of Physics, Damghan University, Damghan, 3671641167, Iran\\
saeed.noorigashti70@gmail.com
\\[2ex]
Behnam Pourhassan \\
School of Physics, Damghan University, Damghan, 3671641167, Iran\\
Center for Theoretical Physics, Khazar University, 41 Mehseti Street,
Baku, AZ1096, Azerbaijan\\
b.pourhassan@du.ac.ir
\\[2ex]
\.{I}zzet Sakall{\i} \\
Physics Department, Eastern Mediterranean University, Famagusta 99628, North Cyprus via Mersin 10, T\"urkiye\\
  izzet.sakalli@emu.edu.tr
}

\begin{document} 

    \maketitle 
\begin{abstract}
WWe study Schwarzschild--AdS black holes in conformal Killing gravity with non-perturbative entropy corrections within the extended phase-space formalism. By deriving the quantum-corrected heat capacity, Helmholtz free energy, internal energy, and Gibbs free energy, we show that quantum corrections modify phase transition points and thermal stability mainly at small horizon radii, while classical behavior is recovered for large black holes. Critical behavior of van der Waals type appears exclusively for positive values of the conformal Killing parameter $a$. Applying the island prescription, we find that the entanglement entropy of Hawking radiation saturates after the Page time at $S_{\text{sat}} = 2(\pi r_H^2 + \eta e^{-\pi r_H^2})$, thereby restoring unitarity. Stronger quantum corrections increase the entropy threshold and delay information recovery. Furthermore, we examine the universal thermodynamic relation in the extremal limit under a minimal perturbation of the AdS curvature radius. We prove that the quantum correction parameter $\eta$ cancels out completely, yielding the robust universal combination $U = -r_{\text{ext}}^3 / l^3$. This demonstrates that the universal relation remains stable against non-perturbative quantum corrections to the entropy.
\end{abstract}

\section{Introduction}
Black holes occupy a unique position in theoretical physics, operating simultaneously as exact solutions to Einstein's field equations and as thermodynamic systems with well-defined temperature and entropy. The foundational realization that black holes behave as macroscopic thermodynamic objects stems from the pioneering work of Bekenstein and Hawking, who demonstrated that entropy is proportional to the horizon area and that emitted radiation possesses a characteristic Hawking temperature \cite{Bekenstein1973,Bekenstein1974,Hawking1975,Bardeen1973,Gibbons1977}. This fundamental link between geometry and thermodynamics, encoded in the Bekenstein--Hawking relation $S_0 = A/4$, remains a cornerstone of modern gravitational physics.

The subsequent formulation of the four laws of black hole mechanics, which map directly onto the laws of classical thermodynamics \cite{Bardeen1973,Wald2001,Sarkar2019}, reinforced the view that black holes can be described by standard thermodynamic variables such as temperature, entropy, pressure, and internal energy.

Extending this framework to asymptotically anti-de Sitter (AdS) spacetimes reveals an exceptionally rich thermodynamic structure. The Hawking--Page transition between thermal radiation and a stable large black hole was an early indicator that AdS black holes exhibit complex thermodynamic behavior \cite{HawkingPage1983,Witten1998,Cai2007,KubiznakMann2015}. Since then, AdS backgrounds have provided an ideal setting to study thermodynamic stability, phase transitions, and the interplay between gravity and quantum mechanics.

In parallel, modified theories of gravity have developed to address open problems in cosmology and quantum gravity, ranging from dark energy to renormalizability
 \cite{Nojiri2011,Mandal2025,Bajardi2024,Cognola2005,deRham2023,Sar}. 
Such modifications significantly alter black hole spacetime geometry and thermodynamics. Notable examples include conformal massive gravity \cite{Soroushfar2021} and various AdS gravity models \cite{Soroushfar2020,Soroushfar2019}. Among these, conformal gravity and conformal Killing gravity are particularly attractive because they yield non-trivial metric modifications while preserving strong geometric principles \cite{Bris}.

Recent studies indicate that modified conformal gravity frameworks induce meaningful changes in horizon structure, geodesic motion, and thermodynamic stability \cite{Soroushfar2019,Soroushfar2021}. In asymptotically AdS spacetimes, these corrections influence both local thermal stability and global phase structure \cite{Soroushfar2020,KubiznakMann2015}, making conformal Killing gravity an excellent laboratory to probe deviations from general relativity.

A second major direction involves quantum corrections to black hole thermodynamics. Near the Planck scale, classical thermodynamics breaks down as quantum gravitational fluctuations modify the entropy--area law. Frameworks such as string theory, loop quantum gravity, and quantum geometry predict both logarithmic perturbative corrections and exponential non-perturbative corrections
\cite{StromingerVafa1996,KaulMajumdar2000,Sen2013,Das2002,BanerjeeMajhi2008,Dab,Mele2024}. 
Non-perturbative terms of the form $\eta e^{-S_0}$ are especially interesting: they are negligible for large black holes but become dominant at small horizon radii, where classical gravity is least reliable. Recent studies have confirmed the importance of such corrections in dirty black holes \cite{Soroushfar2024} and conformally dressed 3D black holes \cite{Soroushfar2023}.

Because black hole entropy reflects both thermodynamic capacity and microscopic quantum information \cite{Bekenstein1973,Hawking1975,Solodukhin2011}, any modification to the entropy--area relation naturally alters the dynamics of information recovery during black hole evaporation \cite{Das2002,Sen2013,Penington2019,Almheiri2020}. This connection directly links quantum-corrected thermodynamics to the black hole information paradox.

In recent years, significant progress has been made toward resolving the information paradox. Hawking's original calculation predicted purely thermal radiation, implying non-unitary evolution \cite{Hawking1975,Hawking1976}. The emergence of the island prescription and quantum extremal surfaces \cite{Penington2019,Almheiri2020} provided a concrete semiclassical mechanism to recover the Page curve \cite{Page1993} and restore unitarity. Further insights from firewall proposals \cite{Almheiri2013}, entanglement wedge reconstruction, and holographic entanglement entropy \cite{Solodukhin2011} have strengthened this picture. The fact that the same corrections affect both thermodynamic potentials and quantum information measures—as seen in Yang--Mills black holes \cite{Soroushfar:2025a}, Rastall--Rainbow compact objects \cite{Eslamzadeh2025}, and brane systems \cite{Pourhassan2022}—points to a unified underlying structure.

In addition to thermal stability and information recovery, another important property of black hole thermodynamics is the universality of extremality relations. In various gravitational setups, specific combinations of thermodynamic variables evaluated in the extremal limit show universal behavior independent of coupling parameters \cite{400,401,402,403,404,405,406}. Testing whether these universal relations remain robust under non-perturbative entropy corrections provides a strong theoretical consistency check.

Despite these advancements, the combined impact of conformal Killing gravity and non-perturbative exponential corrections on thermodynamic stability, information recovery, and universal extremality relations has not yet been explored within a single framework. In this work, we address this problem by introducing the exponential entropy correction
\begin{equation}
S_{\rm corr}=\pi r_H^2+\eta e^{-\pi r_H^2},
\end{equation}
into the spacetime of Schwarzschild--AdS black holes in conformal Killing gravity. We analyze how this modified entropy influences the heat capacity, Helmholtz free energy, internal energy, and Gibbs free energy. Furthermore, using the island prescription, we compute the Page curve and Page time to examine information recovery. Finally, by applying a minimal perturbation to the AdS curvature radius, we evaluate the universal thermodynamic relation in the extremal limit to test its stability against non-perturbative quantum corrections.

The paper is organized as follows. Section~\ref{sec BH} reviews the Schwarzschild--AdS solution in conformal Killing gravity. Section~\ref{sec:entropy} introduces the non-perturbative entropy correction. Thermodynamic properties are analyzed in Section~\ref{sec:thermo}, and information recovery alongside the Page curve is discussed in Section~\ref{sec:page}. In Section~\ref{sec:universal}, we derive and evaluate the universal thermodynamic relation in the extremal limit. Section~\ref{sec con} summarizes our conclusions. Mathematical derivations are detailed in Appendix.
\section{Black Holes in Conformal Killing Gravity}
\label{sec BH}

Conformal Killing gravity is a recently proposed extension of general relativity in which the field equations are constructed from the totally symmetric, traceless tensor $H_{\alpha\mu\nu}$ \cite{Harada2023}. Rather than modifying the action, the theory modifies the field equations directly, and the resulting framework admits black hole solutions whose asymptotic structure differs from that of standard general relativity. We begin by recalling the field equations and the key algebraic properties of the theory, then derive the thermodynamic quantities that will be needed in subsequent sections.

The field equations read \cite{Harada2023,Ladghami2026}
\begin{equation}
H_{\alpha \mu \nu} = 8\pi T_{\alpha \mu \nu},
\label{eq:field_eq}
\end{equation}
where
\begin{equation}
H_{\alpha \mu \nu} = \nabla_{\alpha}R_{\mu \nu} + \nabla_{\mu}R_{\nu \alpha} 
+ \nabla_{\nu}R_{\alpha \mu} 
- \frac{1}{3}\left(g_{\mu \nu}\partial_{\alpha} + g_{\nu \alpha}\partial_{\mu} 
+ g_{\alpha \mu}\partial_{\nu}\right)R,
\label{eq:H_tensor}
\end{equation}
with $R_{\mu\nu}$ the Ricci tensor and $R$ the Ricci scalar. The tensor $H_{\alpha\mu\nu}$ 
is totally symmetric and satisfies the traceless condition
\begin{equation}
g^{\mu \nu}H_{\alpha \mu \nu} = 0.
\label{eq:traceless}
\end{equation}
The generalized energy-momentum tensor $T_{\alpha\mu\nu}$ on the right-hand side is 
defined as
\begin{equation}
T_{\alpha \mu \nu} = \nabla_{\alpha}T_{\mu \nu} + \nabla_{\mu}T_{\nu \alpha} 
+ \nabla_{\nu}T_{\alpha \mu} 
- \frac{1}{6}\left(g_{\mu \nu}\partial_{\alpha} + g_{\nu \alpha}\partial_{\mu} 
+ g_{\alpha \mu}\partial_{\nu}\right)T,
\label{eq:gen_T}
\end{equation}
where $T_{\mu\nu}$ is the standard energy-momentum tensor and $T$ its trace. This tensor  is also totally symmetric and obeys
  \begin{equation}
g^{\mu \nu}T_{\alpha \mu \nu} = 2\nabla_{\mu}T_{\alpha}^{\mu}.
\label{eq:T_trace_relation}
\end{equation}
Contracting the field equation~\eqref{eq:field_eq} with $g^{\mu\nu}$ and using 
Eqs.~\eqref{eq:traceless} and \eqref{eq:T_trace_relation} gives 
$\nabla_{\mu}T^{\mu}_{\alpha} = 0$, so the standard matter conservation law is 
automatically satisfied.

The static, spherically symmetric Schwarzschild--AdS solution in this theory takes the 
form \cite{Ladghami2026}
\begin{equation}
ds^{2} = -f(r_H)\,dt^{2} + \frac{dr_H^{2}}{f(r_H)} + r_H^{2}\left(d\theta^{2} 
+ \sin^{2}\theta\, d\phi^{2}\right),
\label{eq:metric}
\end{equation}
with metric function
\begin{equation}
f(r_H) = 1 - \frac{2M}{r_H} + \frac{r_H^{2}}{l^{2}} - \frac{a}{5}\,r_H^{4}.
\label{eq:metric_function}
\end{equation}
Here $l$ is the AdS curvature radius and $a$ is the conformal Killing gravity 
parameter; the $r_H^4$ term it introduces modifies the large-$r_H$ asymptotics and 
distinguishes this solution from the standard Schwarzschild--AdS geometry, which is 
recovered in the limit $a\to 0$. The mass parameter is
\begin{equation}
M = -\frac{\left(a\,r_H^{4}l^{2}-5l^{2}-5r_H^{2}\right)r_H}{10l^{2}}.
\end{equation}

Imposing the horizon condition $f(r_H)=0$ expresses the mass in terms of the event 
horizon radius,
\begin{equation}
M = \frac{r_{H}}{2}\left(1 + \frac{r_{H}^{2}}{l^{2}} - \frac{a}{5}\,r_{H}^{4}\right).
\label{eq:mass_horizon}
\end{equation}
which reduces to the Schwarzschild--AdS result when $a=0$.

The Hawking temperature follows from the surface gravity $\kappa = f'(r_H)/2$,
\begin{equation}
T_0 = \frac{\kappa}{2\pi} = \frac{1}{4\pi r_{H}}\left(1 + \frac{3r_{H}^{2}}{l^{2}} 
- a\, r_{H}^{4}\right).
\label{eq:hawking_temp}
\end{equation}
Applying the first law  $dM = T_0\,dS_0$ yields the entropy

\begin{equation}
S_0 =  \pi r_{H}^{2},
\label{eq:entropy_BH}
\end{equation}

which satisfies the Bekenstein--Hawking area law. A noteworthy consequence of Eq.~\eqref{eq:entropy_BH} is that the conformal  Killing gravity parameter $a$ is absent from the entropy expression, 
even though the $r_H^4$ correction in the metric function \eqref{eq:metric_function} reshapes the spacetime geometry and shifts the Hawking temperature~\eqref{eq:hawking_temp}. Conformal Killing  Gravity, therefore, leaves its mark on the horizon structure and temperature, but not on the entropy itself. This observation motivates the question of what happens when quantum gravitational effects are taken into account, and it is precisely this question that the remainder of the paper addresses..

\section{Quantum-Corrected Entropy}
\label{sec:entropy}

The Bekenstein--Hawking relation $S_0 = \pi r_H^2$ is a classical result, and there are  strong theoretical reasons to expect deviations from it once quantum gravitational effects  are taken into account, particularly for black holes whose horizon radius approaches the  Planck scale \cite{StromingerVafa1996,Sen2013}. Predictions from string theory and loop  quantum gravity suggest that such deviations fall into two broad classes: perturbative  corrections, most commonly logarithmic in $S_0$, and non-perturbative corrections that  appear as exponential terms in $e^{-S_0}$.  The latter are of particular interest because they are strongly suppressed for macroscopic black holes, where the condition $S_0 \gg 1$ makes the exponential contribution effectively negligible. In contrast, at small horizon radii, the exponential term becomes increasingly significant and may even provide the dominant correction to the entropy. This regime is especially relevant because quantum tunneling effects and strong quantum fluctuations are expected to become important near the final stages of black hole evaporation \cite{Mele2024,Dab}.

In this work we adopt the non-perturbative correction
\begin{equation}
S_{\rm corr} = \pi r_H^2 + \eta\, e^{-\pi r_H^2},
\label{eq:corrected_entropy}
\end{equation}
where $\eta$ is a dimensionless parameter controlling the strength of the quantum  correction \cite{Ladghami2026}. The structure of Eq.~\eqref{eq:corrected_entropy}  ensures that the classical area law is recovered smoothly as $r_H \to \infty$, while 
for small $r_H$ the exponential term grows and substantially modifies the entropy.  Through the thermodynamic relations $T_0 = \partial M/\partial S_0$ and  $C_0= T_0(\partial S_0/\partial T_0)$, this modification propagates into the heat capacity, Helmholtz free energy, and internal energy, whose behavior we analyze in the  following section.
\section{Thermodynamics}
\label{sec:thermo}

In this section, we study the thermodynamic effects of the quantum correction given in Eq.~\eqref{eq:corrected_entropy}. Thermal stability is examined using the heat capacity: configurations with $C>0$ are thermodynamically stable, those with $C<0$ are unstable, and divergences of $C$ indicate second-order phase transitions. We pay special attention to the three main parameters of the model, namely the quantum correction strength $\eta$, the conformal Killing gravity parameter $a$, and the AdS radius $l$, and investigate how they influence the thermodynamic behavior of the black hole.

To see the effect of quantum corrections more clearly, we divide the analysis into three steps. First, we correct only the entropy as in Eq.~\eqref{eq:corrected_entropy}, while keeping the mass and Hawking temperature classical. Second, we use the corrected entropy along with a temperature derived from the corrected first law, but we still keep the mass unchanged. Third, we compare all thermodynamic quantities with the standard Schwarzschild--AdS case. This helps us isolate the effect of quantum corrections and compare the different approaches.
\subsection{Thermodynamics with Corrected Entropy and Classical Temperature}
In this part, we keep the entropy corrected as in Eq.~\eqref{eq:corrected_entropy}, but we use the classical forms for mass and Hawking temperature. This helps us see how the quantum correction alone changes the thermodynamic behavior
\subsubsection{Heat Capacity with Corrected Entropy and Classical Temperature}
\label{subsec:heatcap}

The heat capacity at constant pressure,
\begin{equation}
C_{\rm S} = T_0\left( \frac{\partial S_{\rm corr}}{\partial T_0} \right)_P,
\label{eq:heat_capacity_def}
\end{equation}
governs thermal stability in the extended phase space formalism \cite{Kubiznak2017,Davies1977}: 
configurations with $C_{\rm S} > 0$ are stable against thermal fluctuations, those with $C_{\rm S} < 0$  are not, and divergences in $C_{\rm S}$  where the denominator of  Eq.~\eqref{eq:heat_capacity_def} vanishes  signal second-order phase transitions.

Substituting the corrected entropy $S_{\rm corr} = \pi r_H^2 + \eta e^{-\pi r_H^2}$ and the  temperature~\eqref{eq:hawking_temp} into Eq.~\eqref{eq:heat_capacity_def} and  differentiating yields
\begin{equation}
C_{\rm S} = -\frac{2 r_{H}^{2} \left(a r_{H}^{4} l^{2} - l^{2} - 3 r_{H}^{2}\right) 
\pi \left(\eta e^{-\pi r_{H}^{2}} - 1\right)}{3 a r_{H}^{4} l^{2} + l^{2} - 3 r_{H}^{2}}.
\label{eq:heat_capacity_explicit}
\end{equation}
The denominator $3a r_H^4 l^2 + l^2 - 3r_H^2$ vanishes at the phase transition points. 
The numerator factorizes into two physically distinct contributions: 
$(ar_H^4 l^2 - l^2 - 3r_H^2)$, which encodes the combined effect of conformal gravity 
and the AdS geometry, and $(\eta e^{-\pi r_H^2} - 1)$, which carries the quantum  correction. In the classical limit $\eta = 0$ the second factor reduces to $-1$ and  the standard Schwarzschild--AdS result is recovered; for small $r_H$ the exponential  grows and modifies the thermal behavior precisely where quantum gravitational effects  are expected to be important.

\begin{figure}[h]
  \centering
  \subfigure[$\eta$ dependence]{
    \includegraphics[width=0.45\textwidth]{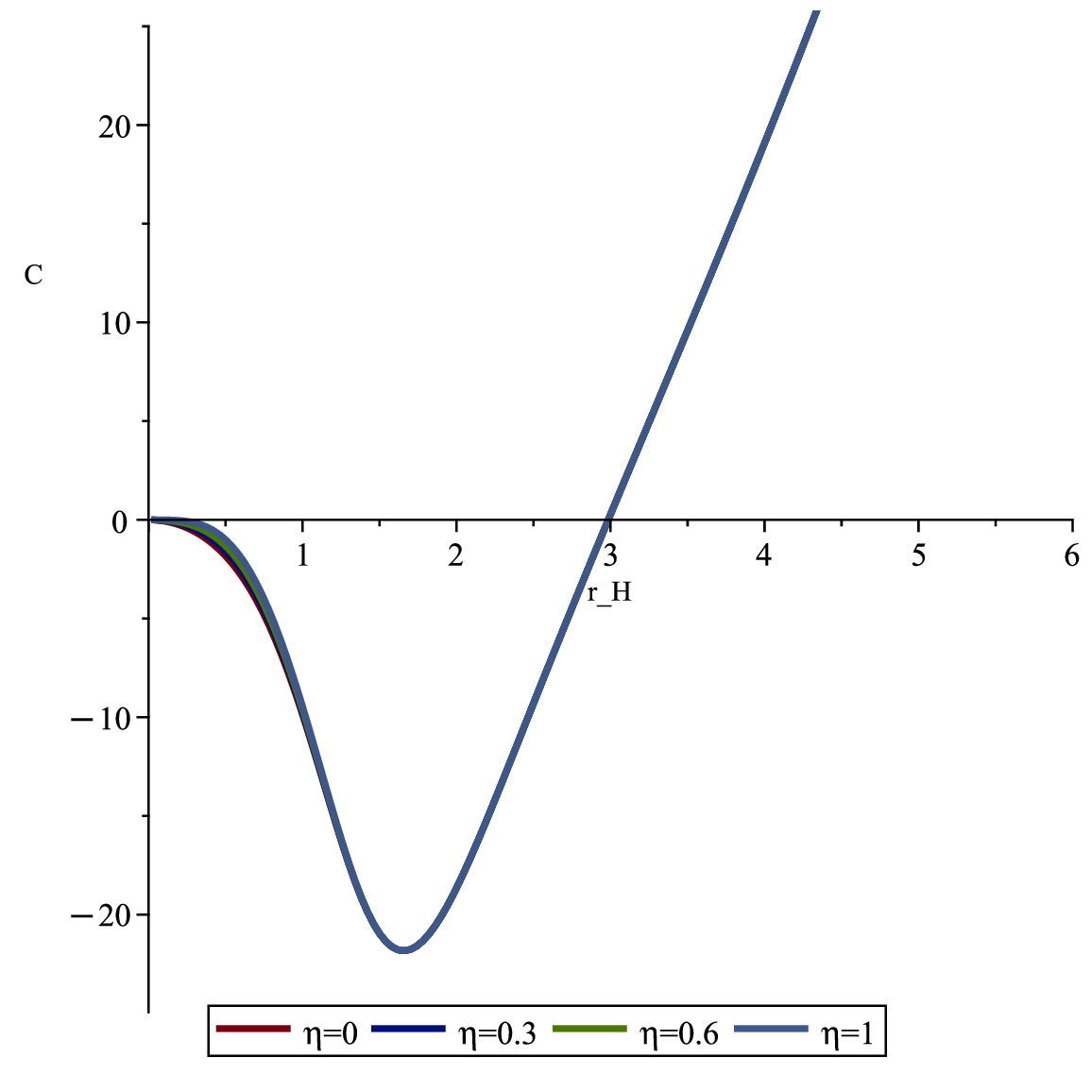}
  }
  \subfigure[Close-up behavior]{
    \includegraphics[width=0.45\textwidth]{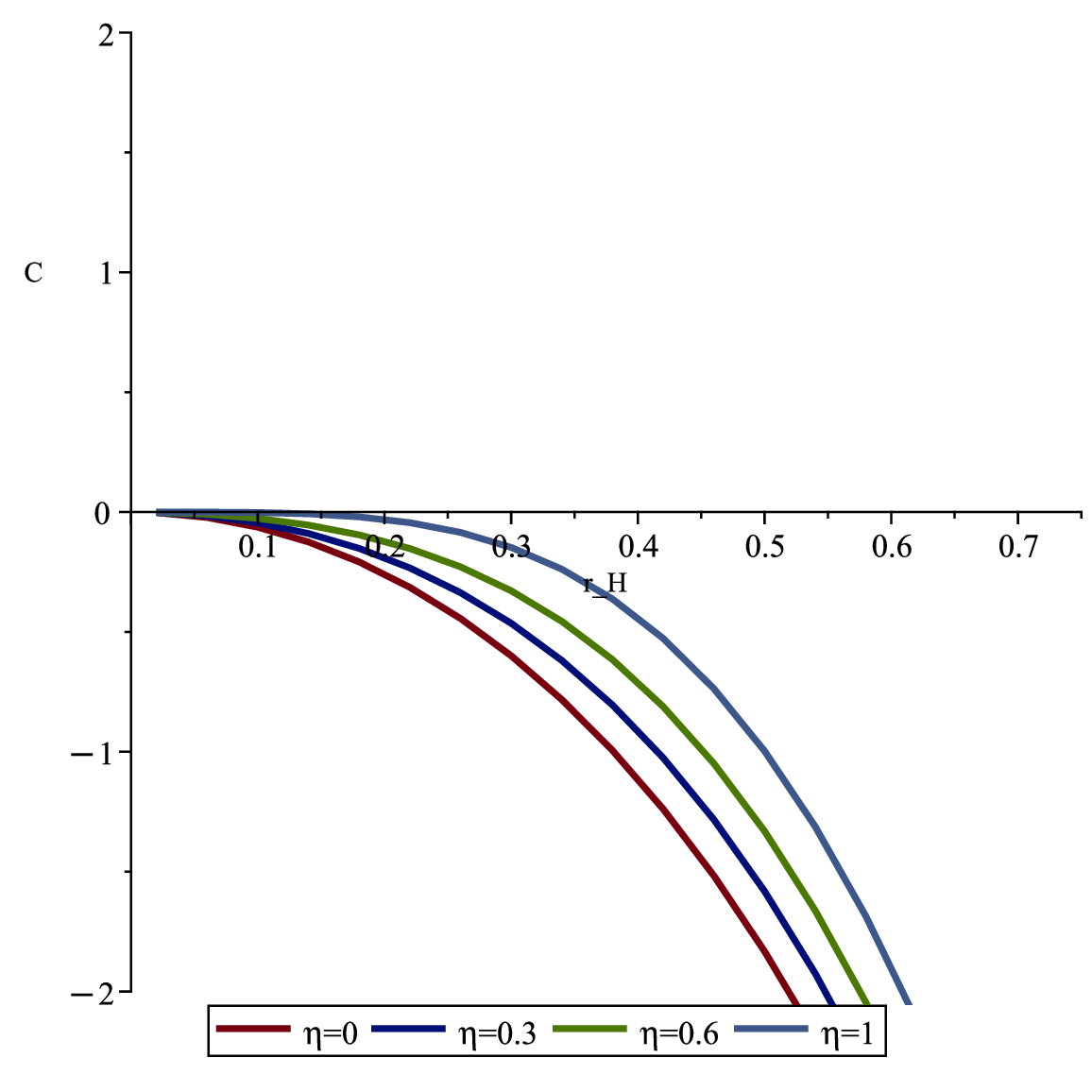}
  }
  \subfigure[$a$ dependence]{
    \includegraphics[width=0.45\textwidth]{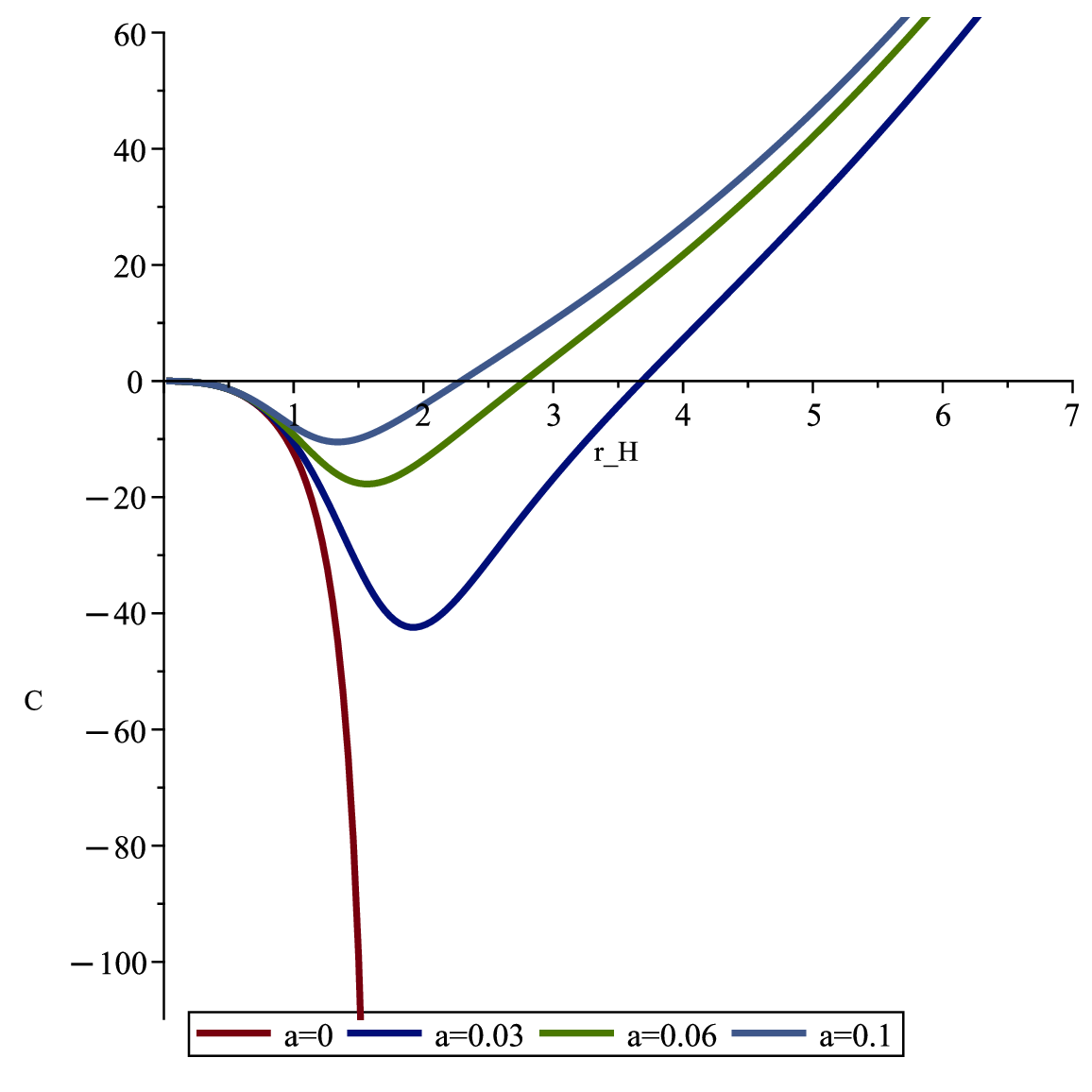}
  }
  \subfigure[$l$ dependence]{
    \includegraphics[width=0.45\textwidth]{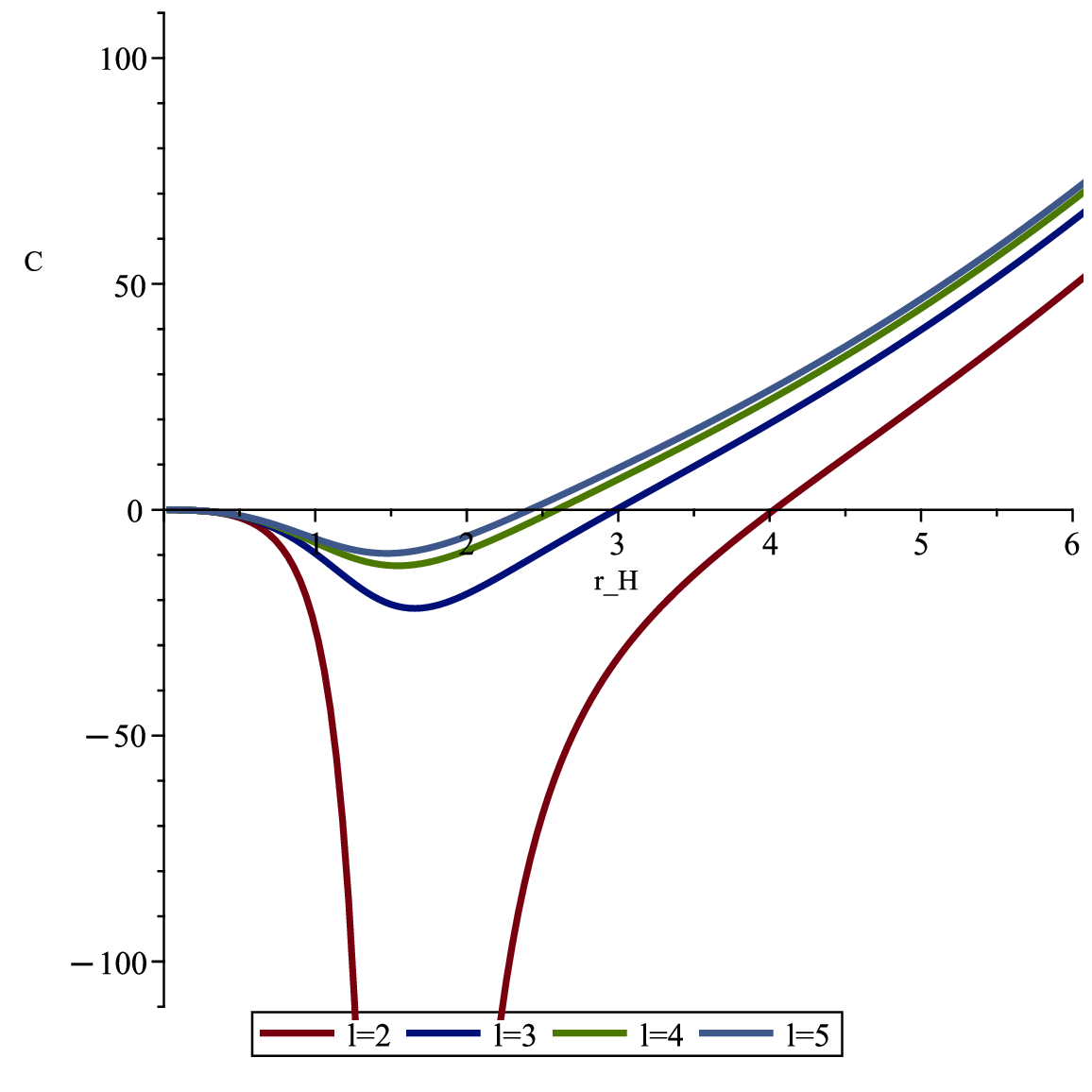}
  }
  \caption{
  Heat capacity of the Schwarzschild AdS black hole in conformal Killing gravity with 
  non-perturbative quantum corrections.
  Panel (a) shows the variation with $\eta$ for fixed $a = 0.05$ and $l = 3$; 
  panel (b) is a close-up of the small-$r_H$ region where quantum effects are most 
  pronounced.
  Panel (c) illustrates the dependence on $a$ for fixed $\eta = 0.5$ and $l = 3$; 
  panel (d) shows the effect of $l$ for fixed $\eta = 0.5$ and $a = 0.05$.
  Divergence points correspond to second-order phase transitions separating stable 
  and unstable phases.
  }
  \label{fig1}
\end{figure}

The dependence of $C_{\rm S}$ on all three parameters is shown in Fig.~\ref{fig1}, with  an additional close-up panel for the small-radius regime.

At $\eta=0$, the heat capacity reduces to the standard Schwarzschild--AdS result and exhibits the familiar divergence associated with a second-order phase transition. Once the non-perturbative correction is introduced, the thermodynamic behavior in the small-horizon-radius region changes noticeably. As shown in panel (a), increasing $\eta$ shifts the location of the divergence and modifies the size of the stable and unstable branches.

The origin of this behavior can be traced directly to the exponential correction factor
$\eta e^{-\pi r_H^2}$ appearing in Eq.~\eqref{eq:heat_capacity_explicit}. For small horizon radii, the exponential term remains finite and contributes significantly to the heat capacity, whereas for large $r_H$ it is exponentially suppressed. Consequently, all curves gradually converge toward the classical Schwarzschild--AdS behavior in the large-radius limit. The enlarged view presented in panel (b) highlights this short-distance quantum regime and clearly shows the departure from the classical profile.

The influence of the conformal Killing gravity parameter $a$ is illustrated in panel (c). Increasing $a$ moves the divergence point toward larger values of the horizon radius and alters the extent of the thermodynamically stable region. Since $a$ enters directly into both the numerator and denominator of Eq.~\eqref{eq:heat_capacity_explicit}, it modifies the balance between the gravitational attraction and the AdS contribution, leading to a noticeable change in the phase structure of the system.

Panel (d) illustrates the influence of the AdS radius $l$. Larger values of $l$ shift the critical radius toward larger horizon sizes and smooth the overall heat-capacity profile. This behavior indicates that the AdS radius significantly affects the thermodynamic structure of the black hole by modifying both the location of the phase transition and the extent of the stable region. Consequently, the AdS background not only influences the global geometry of spacetime but also plays an important role in determining the thermodynamic stability of the black hole.

\subsubsection{Helmholtz Free Energy with Corrected Entropy and Classical Temperature}
\label{subsec:helmholtz}
The Helmholtz free energy determines the global thermodynamic stability of the system. In the canonical ensemble, the preferred equilibrium state is the one that minimizes the free energy, while configurations with higher free energy are thermodynamically disfavored. Therefore, 
the behavior of the Helmholtz free energy tells us about equilibrium configurations and possible phase transitions.

Because the entropy of the present model contains a non-perturbative quantum correction, the Helmholtz free energy must be derived from the same corrected entropy. It is therefore defined through

\begin{equation}
F_{\rm S}=-\int S_{\rm corr}dT_0,
\label{eq:helmholtz_definition}
\end{equation}

where the entropy is given by Eq.~\eqref{eq:corrected_entropy}. Since both the entropy and the Hawking temperature depend on the horizon radius $ (r_H) $, it is convenient to use $ (r_H) $ as the integration variable. Equation~\eqref{eq:helmholtz_definition} can then be written as

\begin{equation}
F_{\rm S}=-\int
S_{\rm corr}(r_H)
\frac{dT_0(r_H)}{dr_H}dr_H .
\label{eq:F_rH}
\end{equation}

Substituting Eqs.~\eqref{eq:corrected_entropy} and \eqref{eq:hawking_temp} into Eq.~\eqref{eq:F_rH} and performing the integration yields

\begin{equation}
F_{\rm S} = \frac{-30 l^{2} \left(a r_{H}^{2} + \frac{2\pi}{3}\right) \eta e^{-\pi r_{H}^{2}} 
+ 12 r_{H} A}{80 \pi^{2} r_{H} l^{2}},
\label{eq:helmholtz_free_energy}
\end{equation}
where
\begin{equation}
A = \frac{5\eta\left(-\frac{4}{3}l^{2}\pi^{2} + al^{2} - 2\pi\right)}{4}\,
\mathrm{erf}\!\left(\sqrt{\pi}\,r_H\right) 
+ \pi^{2} r_{H}\!\left(a r_{H}^{4} l^{2} + \frac{5}{3}l^{2} - \frac{5}{3}r_{H}^{2}\right).
\label{eq:auxiliary_A}
\end{equation}

Here the error function is defined as $\mathrm{erf}(x) = \frac{2}{\sqrt{\pi}}\int_0^x e^{-t^2}dt$, which naturally appears when integrating the exponential term $e^{-\pi r_H^2}$ from the corrected entropy. In the classical limit $ (\eta \rightarrow 0) $, the exponential contribution vanishes and Eq.~\eqref{eq:helmholtz_free_energy} reduces smoothly to the corresponding Schwarzschild--AdS result \cite{Gibbs1876,Callen1985}.

\begin{figure}[h]
\centering
\subfigure[$\eta$ dependence]{
\includegraphics[width=0.45\textwidth]{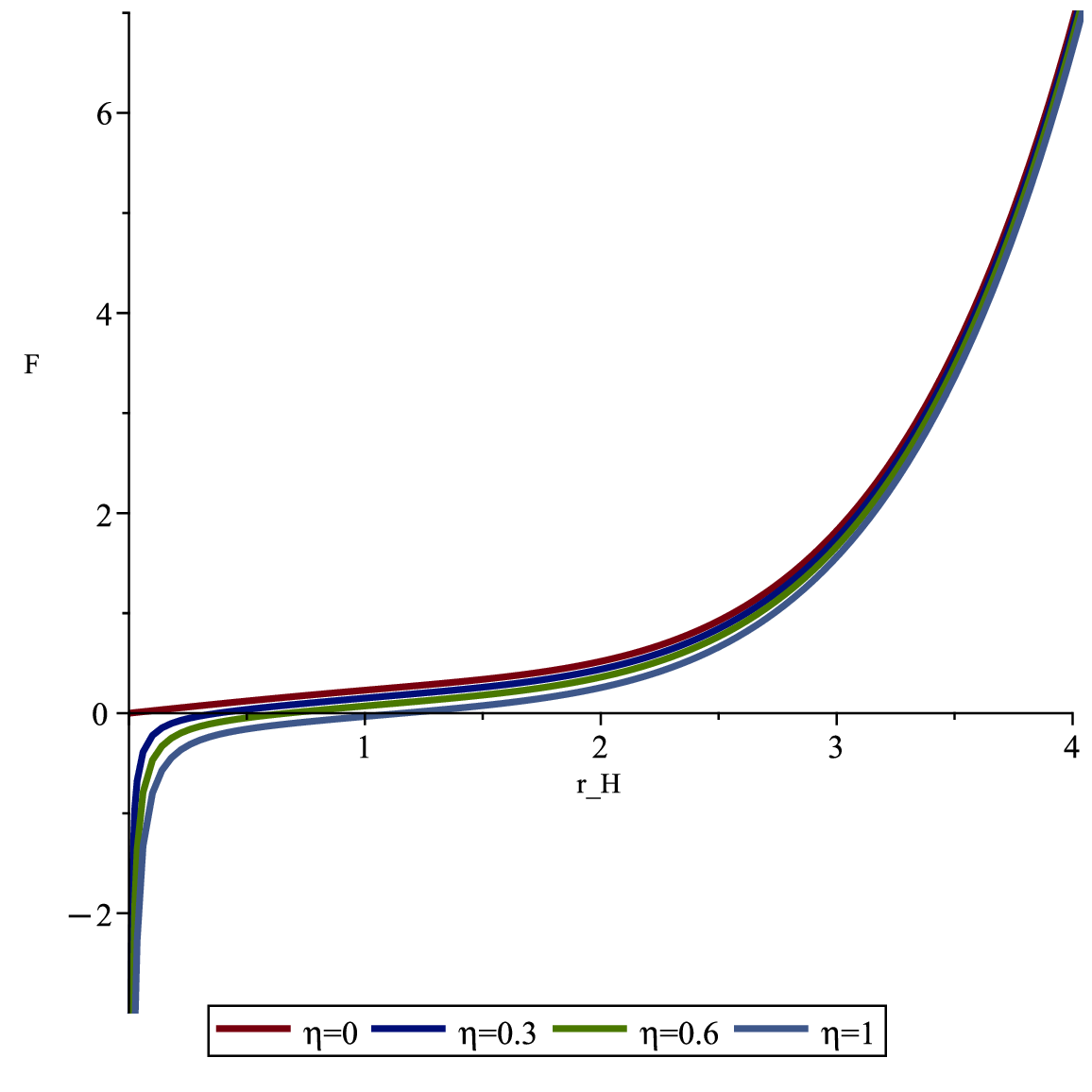}
}
\subfigure[$a$ dependence]{
\includegraphics[width=0.45\textwidth]{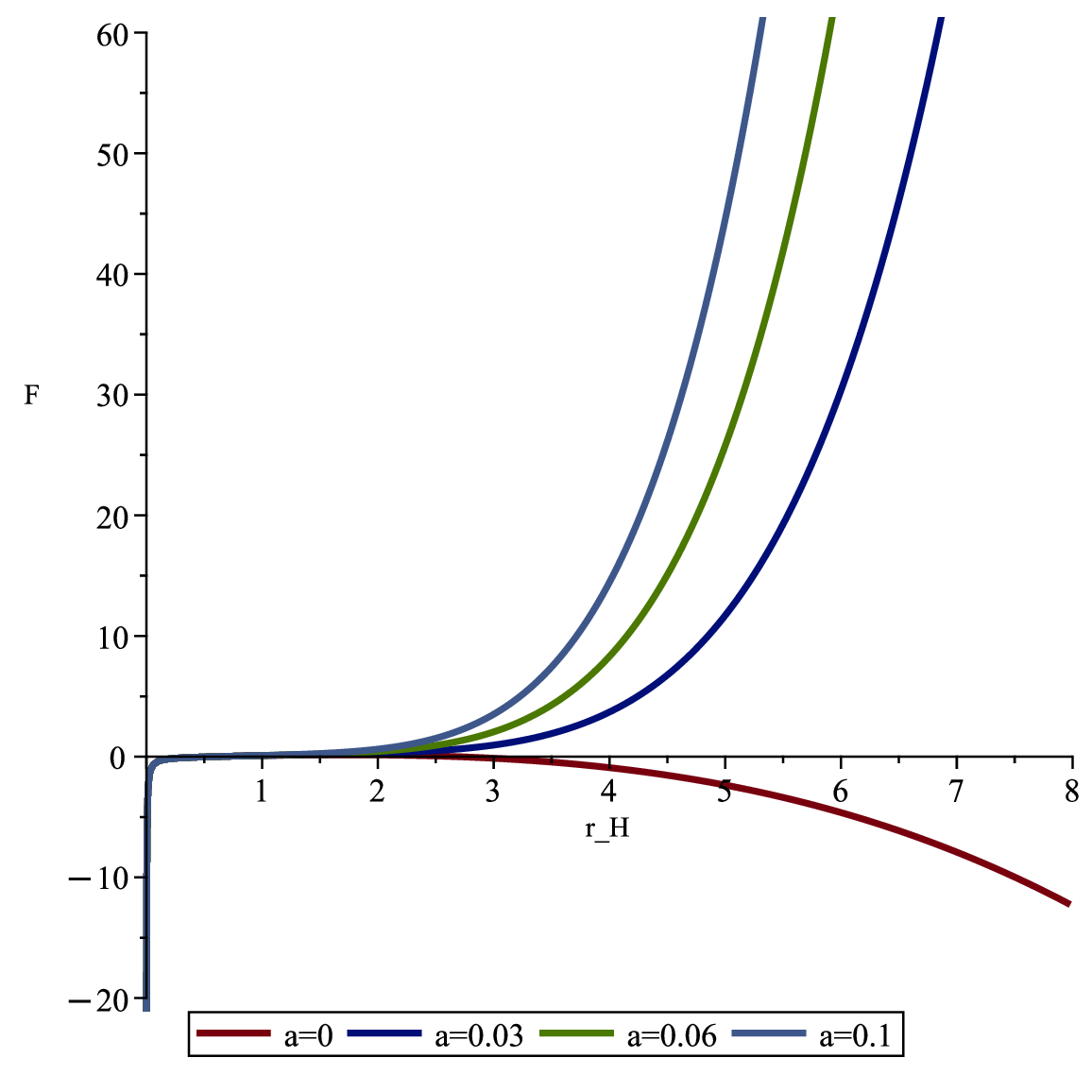}
}
\subfigure[$l$ dependence]{
\includegraphics[width=0.45\textwidth]{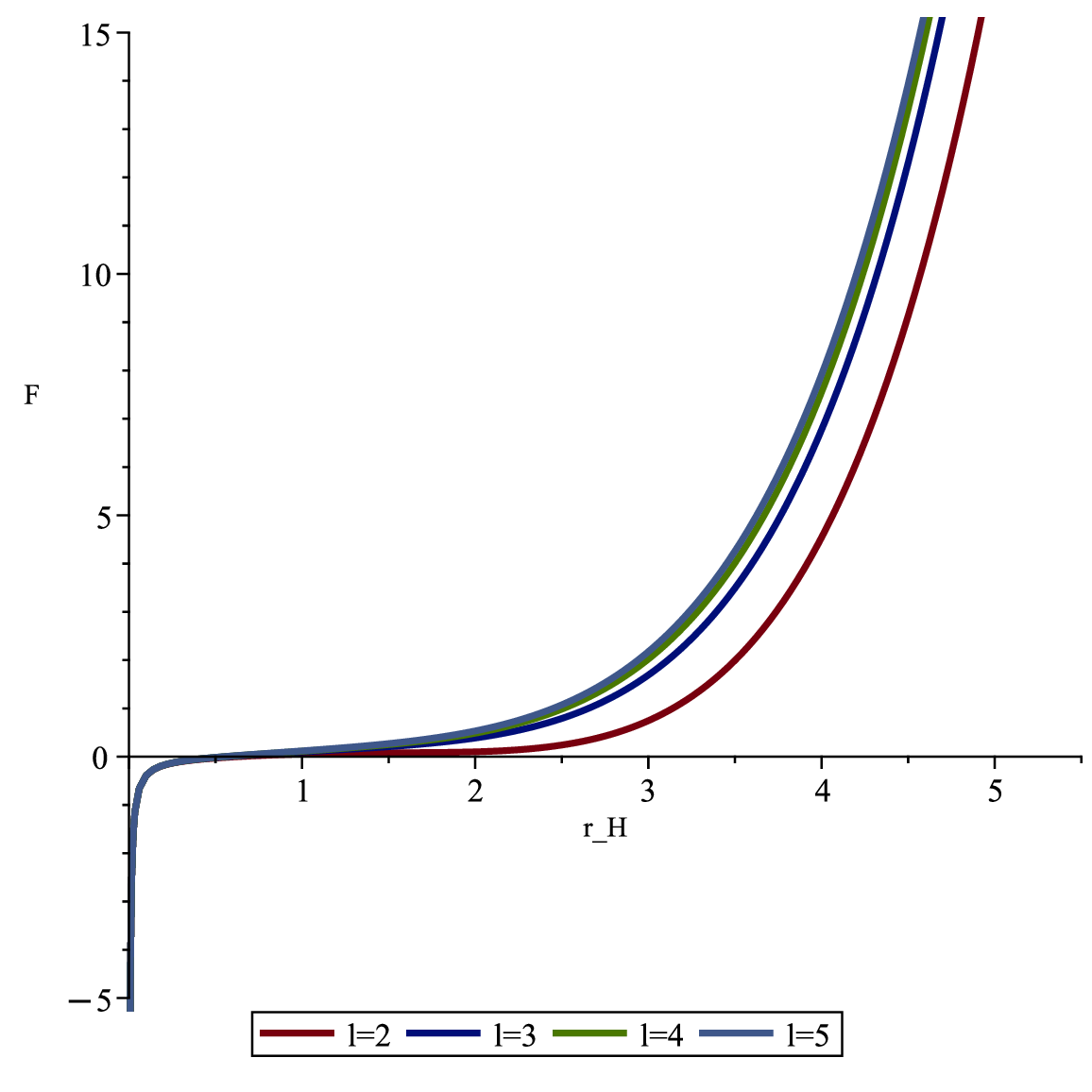}
}
\caption{
Corrected Helmholtz free energy of Schwarzschild AdS black holes in conformal Killing gravity with non-perturbative quantum corrections.
Panel (a) shows the dependence on the quantum correction parameter $\eta$ for fixed $a=0.05$ and $l=3$.
Panel (b) illustrates the effect of the conformal gravity parameter $a$ for fixed $\eta=0.5$ and $l=3$.
Panel (c) presents the variation with the AdS radius $l$ for fixed $\eta=0.5$ and $a=0.05$.
The minima of the free-energy curves correspond to thermodynamically preferred equilibrium configurations.
}
\label{fig2}
\end{figure}

The behavior of the corrected Helmholtz free energy for different values of the quantum correction parameter $\eta$, the conformal gravity parameter $a$, and the AdS radius $l$ is presented in Fig.~\ref{fig2}.

Figure~\ref{fig2} shows that the quantum correction parameter $\eta$ mainly affects the small-horizon-radius regime. As illustrated in panel (a), increasing $\eta$ modifies both the depth and the location of the free-energy minimum, indicating that quantum effects alter the preferred equilibrium configuration of the black hole. These deviations become significant only in the quantum regime, while for large horizon radius all curves approach the same asymptotic behavior.

The influence of the conformal gravity parameter $a$ is displayed in panel (b). Increasing $a$ shifts the minimum of the free energy toward larger horizon radius and modifies the overall shape of the thermodynamic profile. Unlike the quantum correction parameter, whose effects are localized near the microscopic regime, the contribution of $a$ remains important over a wider range of horizon radii and therefore affects the global thermodynamic structure.

In contrast to $a=0$, where the free energy in panel (b) remains positive and grows steadily with $r_H$, the curves with  $ a>0 $
 turn negative at large $r_H$, with the downturn occurring at progressively smaller $r_H$ as a increases. This indicates that switching on the conformal Killing gravity correction qualitatively changes the large-$r_H$ behavior of the free energy relative to the classical Schwarzschild–AdS case: the term
 $ar_H^4 l^2$ Eq.~\eqref{eq:auxiliary_A}, which is absent when $ a=0 $and grows as  $r_H^4$ for $ a>0 $, drives the free energy negative once $r_H$ is large
  enough. This is not a breakdown of the formalism, but it does mean that for $ a>0 $, the free-energy minimum and the large-$r_H$ negative branch coexist, and their relation to the stability conditions identified through the heat capacity in Sec.~\ref{subsec:heatcap} deserves closer examination.

Panel (c) demonstrates the role of the AdS radius $l$. Larger values of $l$ smooth the free-energy profile and shift the equilibrium point toward larger horizon radii. This behavior shows that the AdS radius has a significant influence on the global thermodynamic equilibrium of the black hole by modifying the shape of the free-energy landscape and the location of the preferred equilibrium configuration. Consequently, the AdS background not only affects the spacetime geometry but also plays an important role in determining the thermodynamic behavior of the system.

Overall, the corrected Helmholtz free energy confirms that the combined effects of conformal Killing gravity and non-perturbative quantum corrections generate a richer thermodynamic structure than that of the classical Schwarzschild--AdS black hole. While quantum corrections dominate the microscopic regime, the parameters $a$ and $l$ govern the large-scale thermodynamic behavior and equilibrium structure.
\subsubsection{Internal Energy with Corrected Entropy and Classical Temperature}
\label{subsec:internal}

The internal energy provides a measure of the total thermodynamic energy stored in the black-hole system and plays an important role in understanding the energetic consequences of quantum corrections. Since the entropy of the present model receives non-perturbative modifications, thermodynamic consistency requires that the internal energy be derived from the same corrected entropy.

Accordingly, the corrected internal energy is obtained by integrating the first law of thermodynamics at constant volume \cite{Bose1997,Callen1985},

\begin{equation}
E_{\rm S}
 = 
\int T_0dS_{\rm corr}
 = 
\int T_0(r_H)\frac{dS_{\rm corr}(r_H)}{dr_H}dr_H,
\label{eq:internal_energy_from_entropy}
\end{equation}
where $S_{\rm corr} = \pi r_H^2 + \eta e^{-\pi r_H^2}$
is the corrected entropy given in Eq.~\eqref{eq:corrected_entropy}. The integration constant is fixed by the choice of the reference background. This construction guarantees that the internal energy, entropy, and Helmholtz free energy all originate from a common thermodynamic framework and satisfy the standard thermodynamic relations consistently.

Changing variables from $S_{\rm corr}$ to the horizon radius $r_H$ and performing the integration analytically yields
\begin{equation}
E_{\rm S} = \frac{-20 r_{H} \left[\left(a l^{2} r_{H}^{2} - 3\right)\pi 
+ \frac{3al^{2}}{2}\right] \eta e^{-\pi r_{H}^{2}} + B}{80\pi^{2} l^{2}},
\label{eq:internal_energy}
\end{equation}
where, 
\begin{equation}
B = 15\eta\!\left(-\frac{4}{3}\pi^{2}l^{2} + al^{2} - 2\pi\right)
\mathrm{erf}\!\left(\sqrt{\pi}\,r_H\right) 
- 8\pi^{2}r_H\!\left[\left(ar_H^{4} - 5\right)l^{2} - 5r_H^{2}\right].
\label{eq:auxiliary_B}
\end{equation}
as in the Helmholtz case, the integration of the exponential correction generates 
an error function.

As in the Helmholtz free energy, the appearance of the error function originates from the integration of the exponential non-perturbative correction term. In the classical limit $\eta\rightarrow0$, all quantum contributions disappear and the standard Schwarzschild--AdS expression is recovered.

\begin{figure}[h]
\centering
\subfigure[$\eta$ dependence]{
\includegraphics[width=0.45\textwidth]{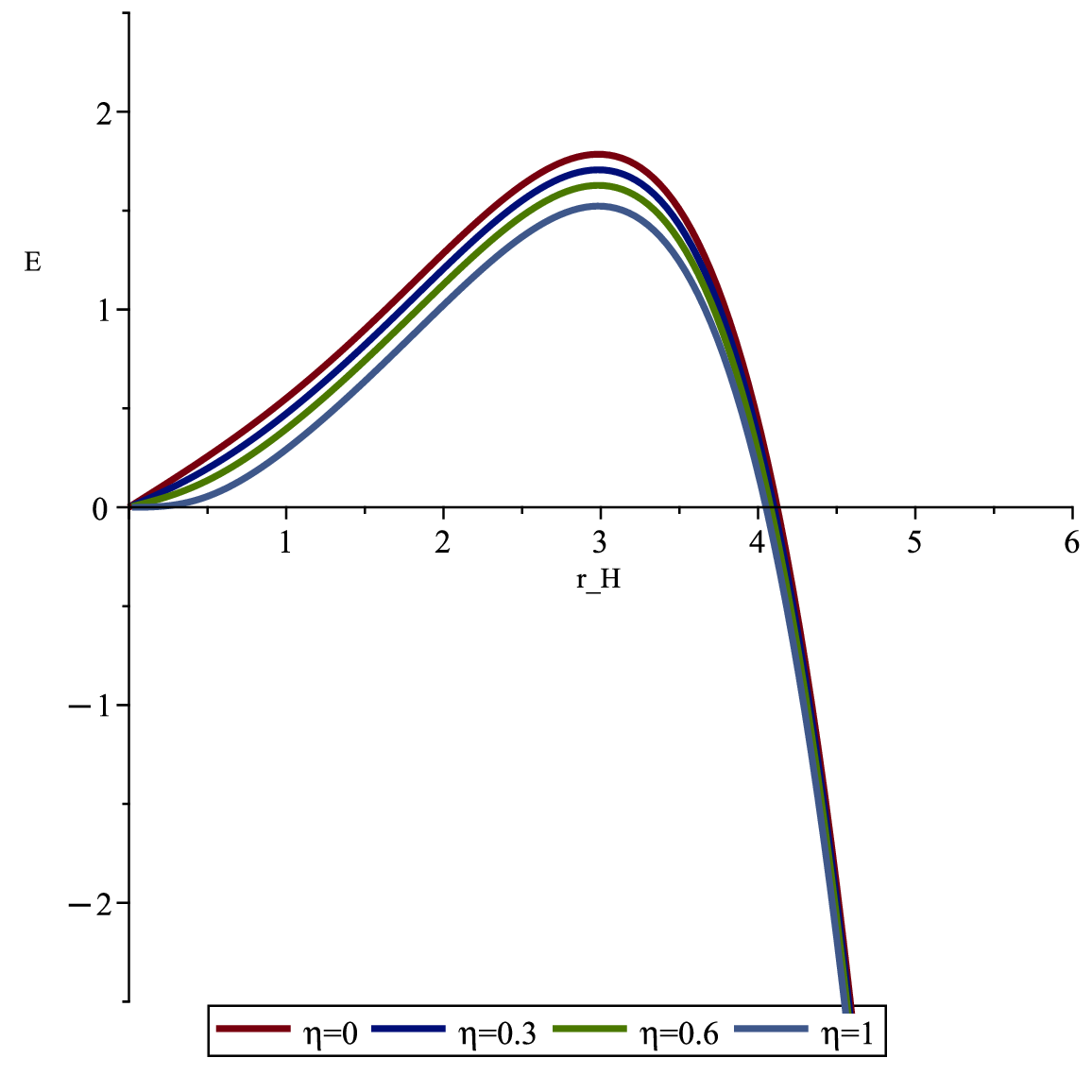}
}
\subfigure[$a$ dependence]{
\includegraphics[width=0.45\textwidth]{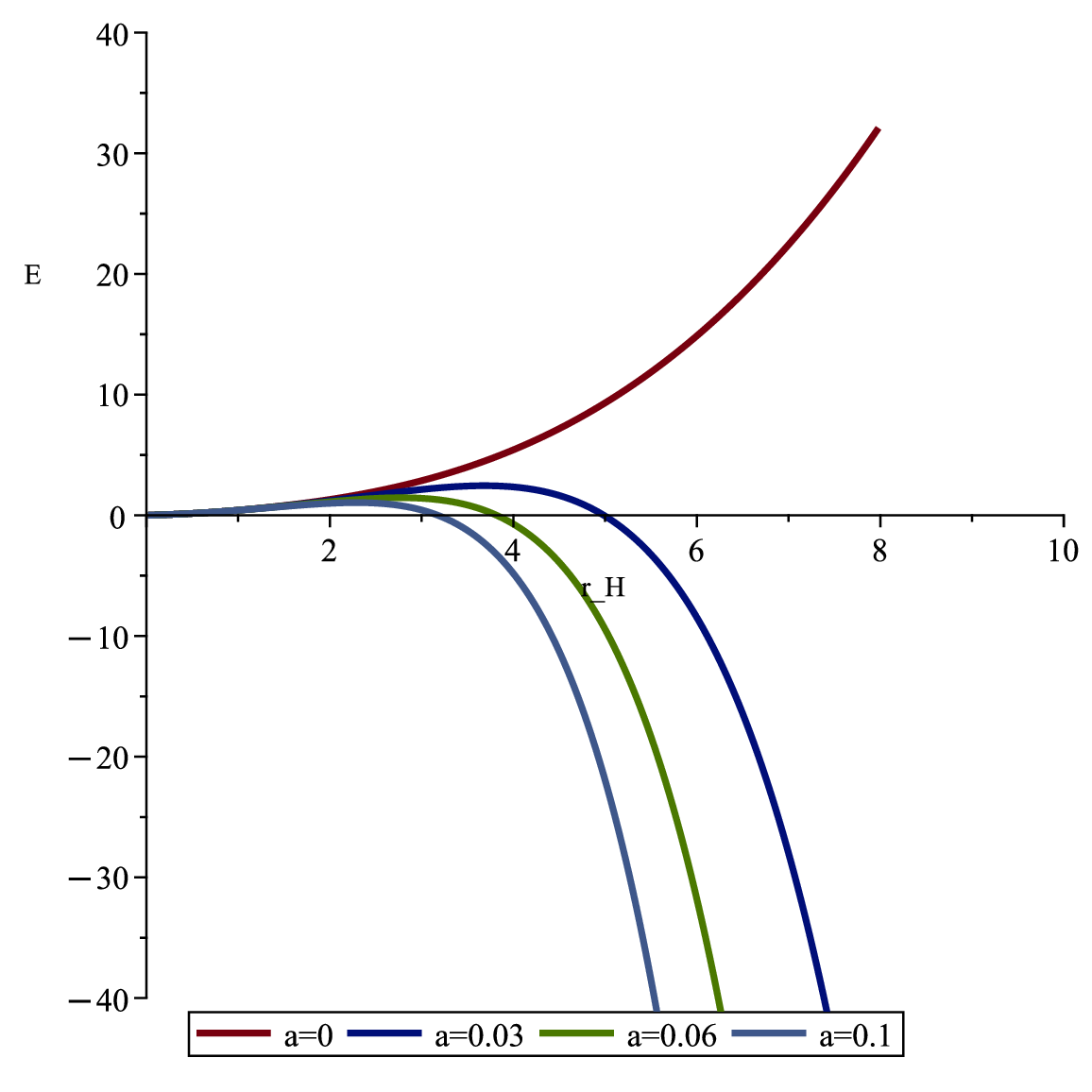}
}
\subfigure[$l$ dependence]{
\includegraphics[width=0.45\textwidth]{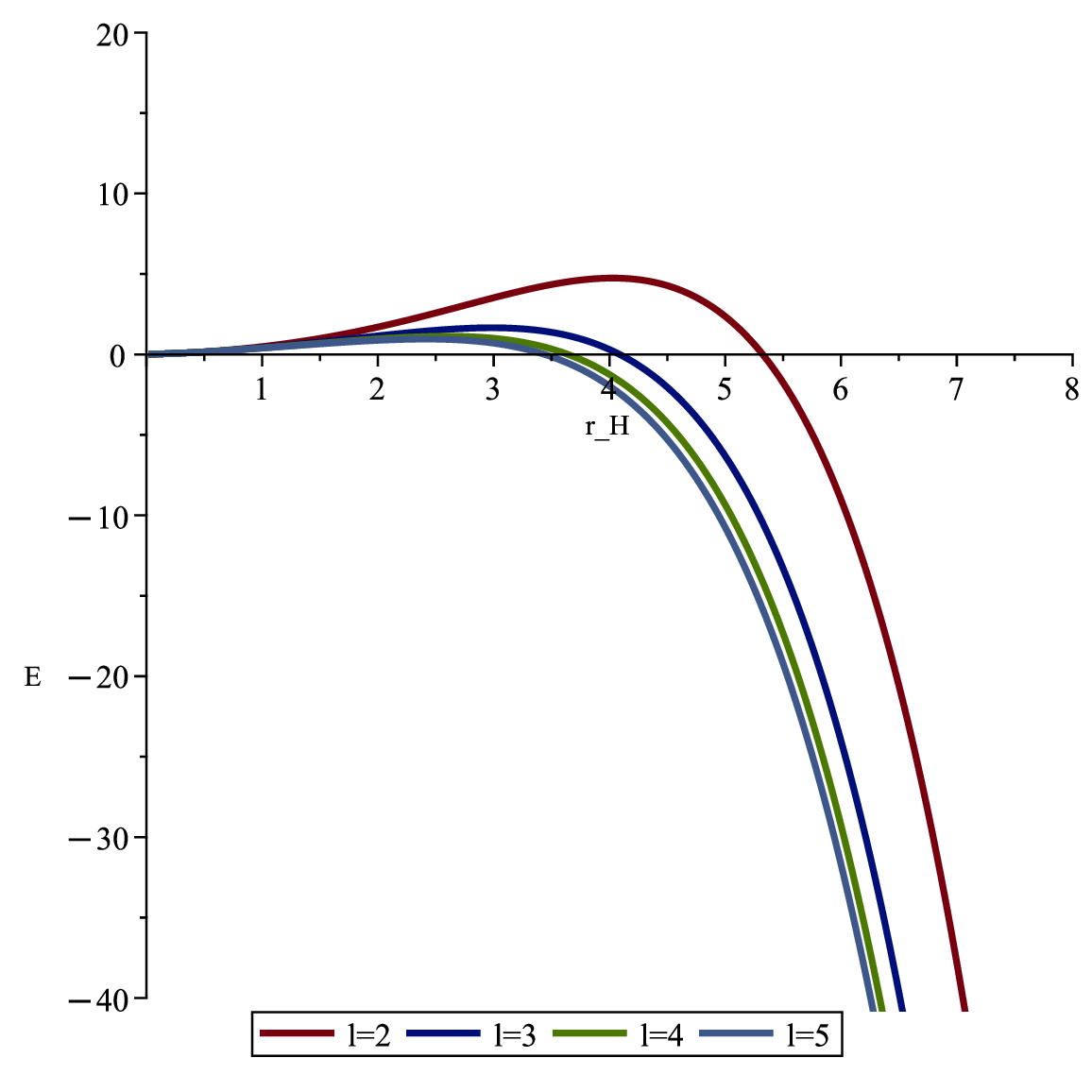}
}
\caption{
Corrected internal energy of Schwarzschild AdS black holes in conformal Killing gravity with non-perturbative quantum corrections.
Panel (a) shows the dependence on the quantum correction parameter $\eta$ for fixed $a=0.05$ and $l=3$.
Panel (b) illustrates the variation with the conformal gravity parameter $a$ for fixed $\eta=0.5$ and $l=3$.
Panel (c) presents the effect of the AdS radius $l$ for fixed $\eta=0.5$ and $a=0.05$.
}
\label{fig3}
\end{figure}

The behavior of the corrected internal energy for different values of the model parameters is displayed in Fig.~\ref{fig3}.

Figure~\ref{fig3} demonstrates that the corrected internal energy grows with the horizon radius in the small- and intermediate-$r_H$ regime, reflecting the growth of the black-hole mass and the corresponding increase in the total energy content of the system, before eventually decreasing and turning negative at large $r_H$ for large enough $a$ or small enough $l$, as discussed below.

The influence of the quantum correction parameter $\eta$ is shown in panel (a). Increasing $\eta$ modifies the internal energy primarily in the small-horizon-radius regime, where quantum effects become important. The deviations rapidly decrease with increasing $r_H$, confirming that the non-perturbative correction acts predominantly at short distances.

Panel (b) illustrates the effect of the conformal gravity parameter $a$. Larger values of $a$ increase the growth rate of the internal energy and modify its asymptotic behavior, indicating that conformal Killing gravity contributes significantly to the large-scale thermodynamic structure.

An important feature visible in panels (b) and (c) is that $E_{\rm corr}$ becomes negative once $r_H$ is large enough. This should not be read as a problem with the model. In the extended phase space formalism $M$ plays the role of enthalpy, not internal energy in the usual sense, and $E_{\rm corr}$ as defined in Eq.~\eqref{eq:internal_energy_from_entropy} comes only from integrating $T_0dS_{\rm corr}$, with no contribution from the $VdP$ and $Ada$ terms that also appear in $dM$. So the sign of $E_{\rm corr}$ does not tell us anything about the thermodynamic stability of the black hole; stability is governed by the heat capacity discussed in Sec.~\ref{subsec:heatcap} and the free energy discussed in Sec.~\ref{subsec:helmholtz}. The sign change instead reflects the growing weight of the $a$ and $l$ terms in Eq.~\eqref{eq:auxiliary_B} at large $r_H$, which enter with negative coefficients in this regime.

The role of the AdS radius $l$ is presented in panel (c). Increasing $l$ smooths the energy profile and changes the overall energy scale of the system. Since $l$ is directly related to the cosmological constant and the thermodynamic pressure, 
this behavior reflects the influence of the background geometry on the energetic properties of the black hole.

Overall, the corrected internal energy confirms that quantum corrections dominate the microscopic regime, while the parameters associated with conformal gravity and AdS geometry govern the macroscopic thermodynamic behavior of the system. The interplay of these contributions generates a richer energy structure than that of the classical Schwarzschild--AdS black hole.
\subsubsection{Gibbs Free Energy with Corrected Entropy and Classical Temperature}
\label{subsec:gibbs_S}

The Gibbs free energy is the potential that determines the globally preferred equilibrium state of a thermodynamic system. In the extended phase-space
formalism, where the black-hole mass \(M\) plays the role of enthalpy, the Gibbs free energy is defined as

\begin{equation}
G_S = M - T_0 S_{\rm corr},
\label{eq:GS_def}
\end{equation}

where \(T_0\) is the classical Hawking temperature
Eq.~\eqref{eq:hawking_temp}
and
\(S_{\rm corr}\) is the corrected entropy  Eq.~\eqref{eq:corrected_entropy}.

The physical meaning of \(G_S\) is transparent: configurations with \(G_S < 0\) correspond to a thermodynamically stable black-hole phase,
while \(G_S > 0\) favours thermal radiation. The point \(G_S = 0\) marks the Hawking Page transition temperature \(T_{\rm HP}\), where the black hole and thermal radiation are in equilibrium.

Substituting the explicit forms of \(M(r_H)\), \(T_0(r_H)\), and
\(S_{\rm corr}(r_H)\) gives

\begin{equation}
G_S(r_H) = \frac{r_H}{4} - \frac{2}{3}\pi P r_H^3 + \frac{3a}{20} r_H^5
- \eta e^{-\pi r_H^2} \frac{8\pi P r_H^2 - a r_H^4 + 1}{4\pi r_H}.
\label{eq:GS_explicit}
\end{equation}

\begin{figure}[h]
\centering
\subfigure[$l$ dependence]{
\includegraphics[width=0.32\textwidth]{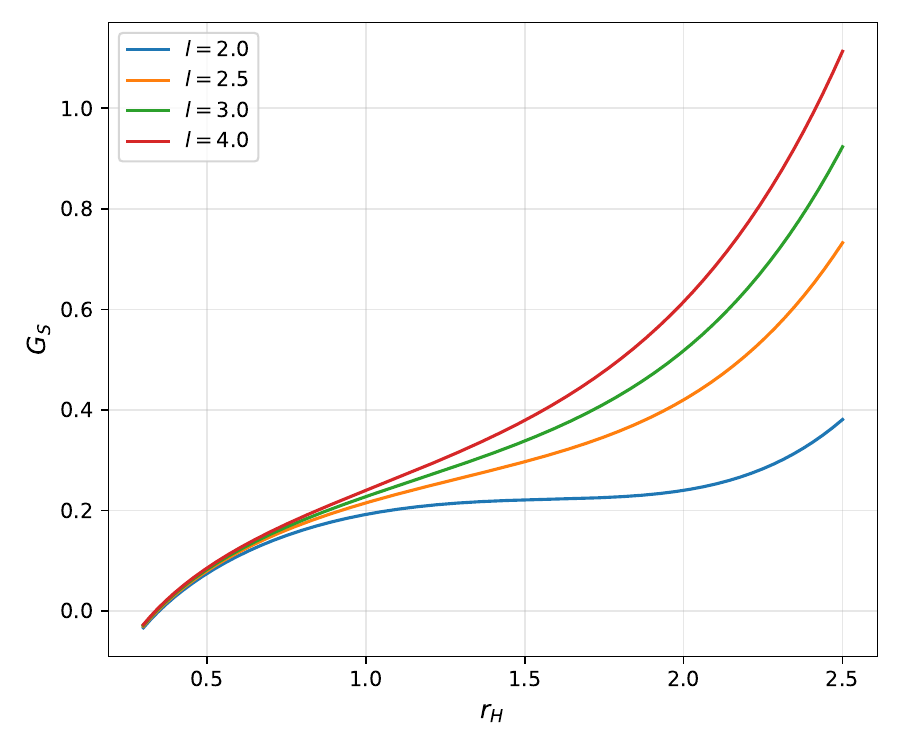}
}
\subfigure[$\eta$ dependence]{
\includegraphics[width=0.32\textwidth]{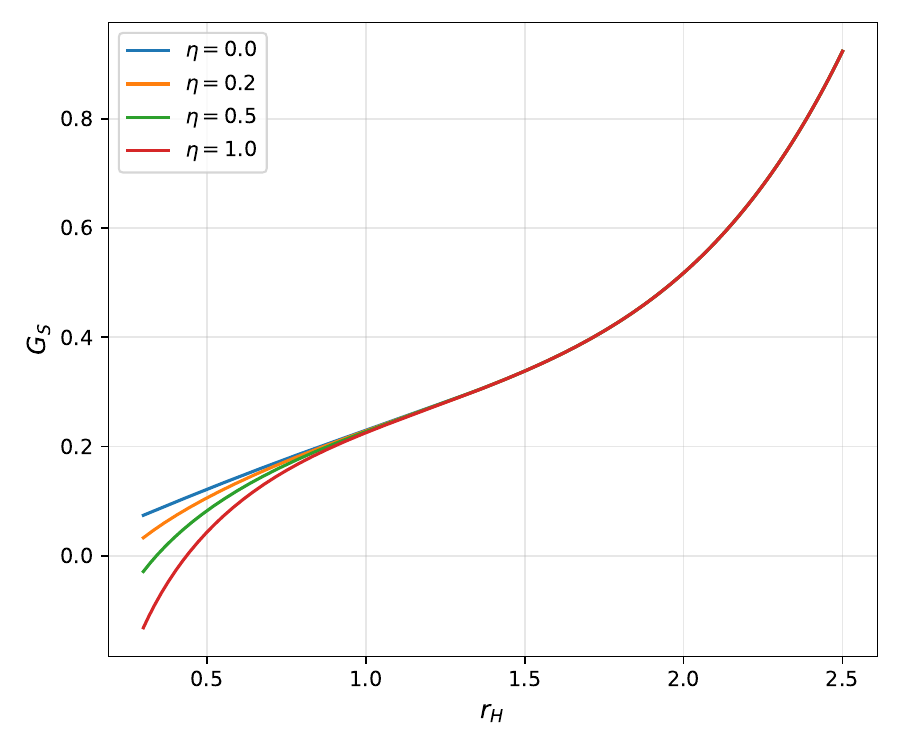}
}
\subfigure[$a$ dependence]{
\includegraphics[width=0.32\textwidth]{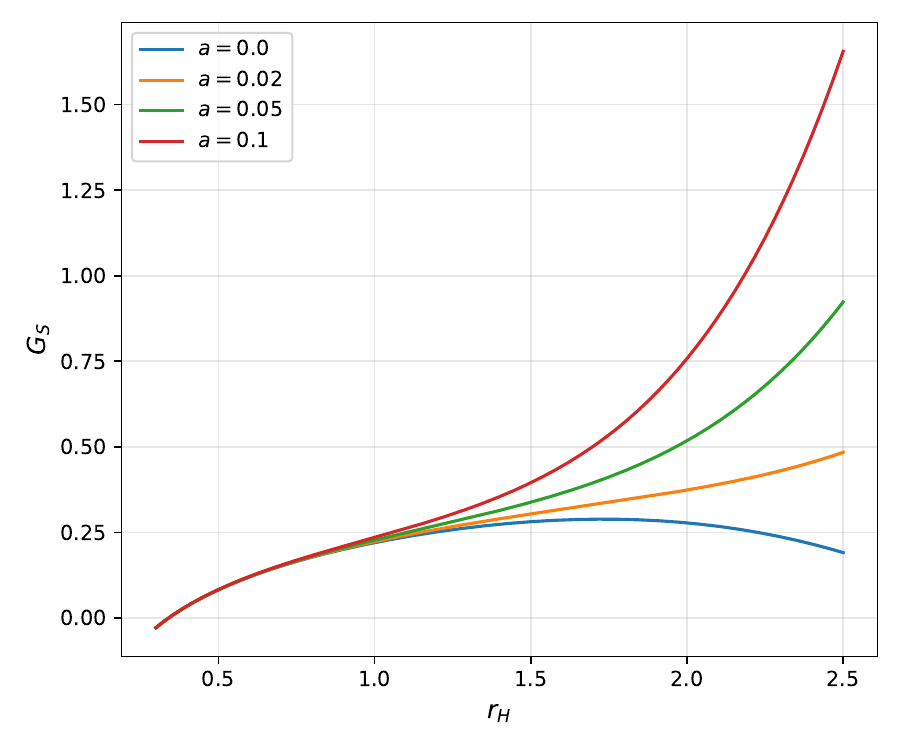}
}
\caption{
Gibbs free energy \(G_S = M - T_0 S_{\rm corr}\) of the Schwarzschild--AdS
black hole in conformal Killing gravity with non-perturbative quantum
corrections.
Panel (a) shows the variation with the AdS radius \(l\) for fixed
\(\eta = 0.5\) and \(a = 0.05\); increasing \(l\) smooths the profile and
shifts the minimum to larger radii. Panel (b) illustrates the dependence on the quantum correction parameter
\(\eta\) for fixed \(a = 0.05\) and \(l = 3\); the curves are nearly indistinguishable for large \(r_H\), with small deviations only in the
small-radius regime.
Panel (c) presents the effect of the conformal Killing gravity parameter \(a\) for fixed \(\eta = 0.5\) and \(l = 3\); increasing \(a\) deepens the
minimum and shifts it to smaller radii.
The crossings \(G_S = 0\) indicate the Hawking Page transition temperature
\(T_{\rm HP}\).
}
\label{fig4}
\end{figure}

The dependence of \(G_S\) on the three fundamental parameters of the model is
shown in Fig.~\ref{fig4}.

Figure~\ref{fig4} shows that the quantum correction parameter \(\eta\) has
only a mild effect on \(G_S\), and only in the small-horizon-radius regime. This is because \(\eta\) enters \(G_S\) only through the entropy correction
\(\eta e^{-\pi r_H^2}\), which decays exponentially. For large \(r_H\), the curves converge to the classical behaviour.

In contrast, the conformal Killing gravity parameter \(a\) has a much more
pronounced effect. For \(a = 0\), the Gibbs free energy remains positive and increases with \(r_H\). Once \(a > 0\), \(G_S\) develops a minimum and
then becomes negative at large radii. This qualitative change reflects the \(a r_H^4\) term in the metric function, which modifies the asymptotic behaviour of the system. Larger values of \(a\) deepen the minimum and
shift it to smaller \(r_H\).

The AdS radius \(l\) also influences \(G_S\): larger \(l\) (lower pressure) smooths the free-energy profile and moves the minimum toward larger horizon
radii. This behaviour is consistent with the identification
\(P = 3/(8\pi l^2)\).

The fact that the quantum correction is most important for small black holes, while the conformal gravity and AdS parameters affect the global
thermodynamic structure, highlights the different roles these parameters play in the model.

\begin{figure}[h]
\centering
\includegraphics[width=0.55\textwidth]{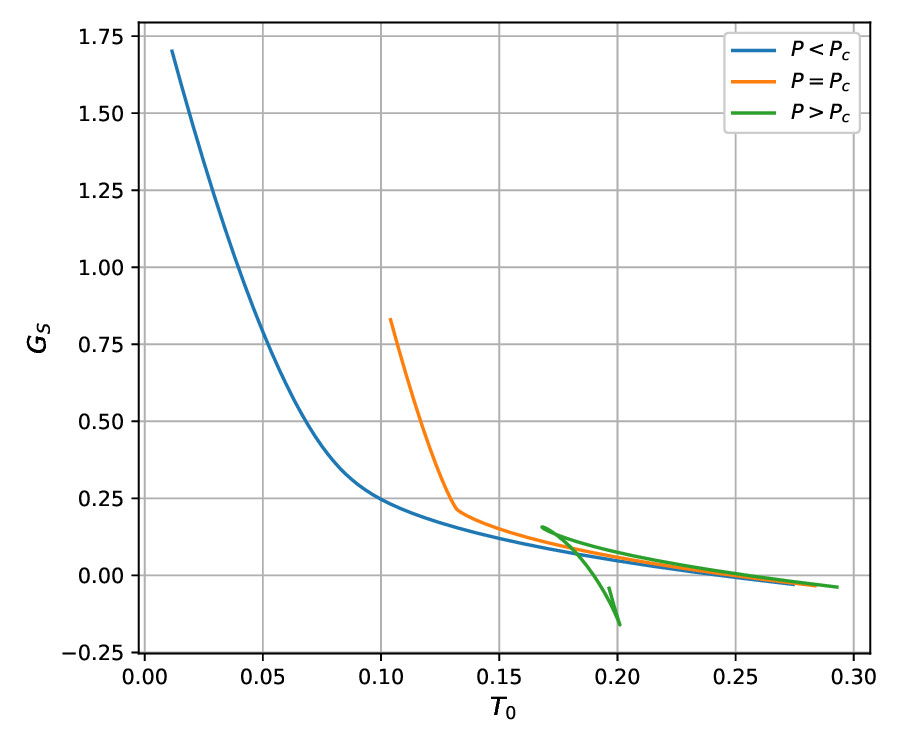}
\caption{
Gibbs free energy \(G_S\) as a function of the classical temperature \(T_0\)
for three values of the pressure \(P\): \(P < P_c\) (solid), \(P = P_c\)
(dashed), and \(P > P_c\) (dotted), with \(a = 0.05\) and \(\eta = 0.5\)
fixed.
For \(P < P_c\), the curve exhibits a swallowtail structure, signalling a
first-order phase transition. At \(P = P_c\), the swallowtail shrinks to a
single point, marking the critical point. For \(P > P_c\), the curve is
monotonic and no phase transition occurs.
}
\label{fig5}
\end{figure}

The behaviour of the Gibbs free energy as a function of temperature provides a direct way to identify the thermodynamically preferred phase and to locate the Hawking Page transition. In Fig.~\ref{fig5}, we plot \(G_S\) against the classical temperature \(T_0\) for three different values of the pressure: below, at, and above the critical pressure \(P_c\).

Figure~\ref{fig5} shows that for pressures below the critical value \(P < P_c\), the Gibbs free energy develops a characteristic swallowtail structure. This behaviour is the signature of a first-order phase transition between small and large black holes. As the pressure increases towards \(P_c\), the swallowtail shrinks, and at the critical pressure \(P = P_c\), it reduces to a single inflection point. Above the
critical pressure, \(P > P_c\), the swallowtail disappears completely and the free energy becomes a monotonic function of temperature, indicating that
no phase transition occurs.

This is a standard feature of van der Waals-type behaviour in black hole thermodynamics, and it confirms that the conformal Killing gravity
parameter \(a > 0\) is responsible for the existence of critical behaviour. The quantum correction \(\eta\) enters through the entropy and affects the
details of the curves, but the overall structure is governed by the pressure relative to \(P_c\).

The equation of state relates the pressure \(P\) to the horizon radius \(r_H\) and the temperature \(T_0\). It is obtained by solving the classical
temperature expression Eq.~\eqref{eq:hawking_temp} for \(P\):

\begin{equation}
P = \frac{T_0}{2r_H} - \frac{1 - a r_H^4}{8\pi r_H^2}.
\label{eq:EOS_S}
\end{equation}

Because the classical temperature \(T_0\) is used, the quantum correction parameter \(\eta\) does not appear explicitly in this equation. It affects
the thermodynamics only through the entropy and the derived quantities such
as \(G_S\).

\begin{figure}[h]
\centering
\subfigure[Isotherms \(T_0\)]{
\includegraphics[width=0.45\textwidth]{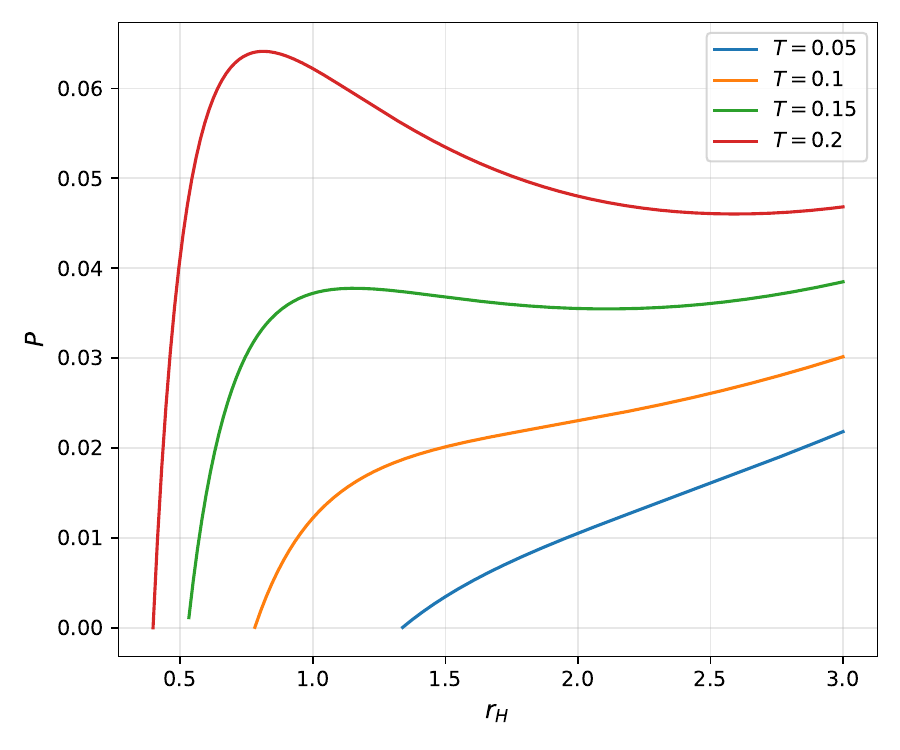}
}
\subfigure[$a$ dependence]{
\includegraphics[width=0.45\textwidth]{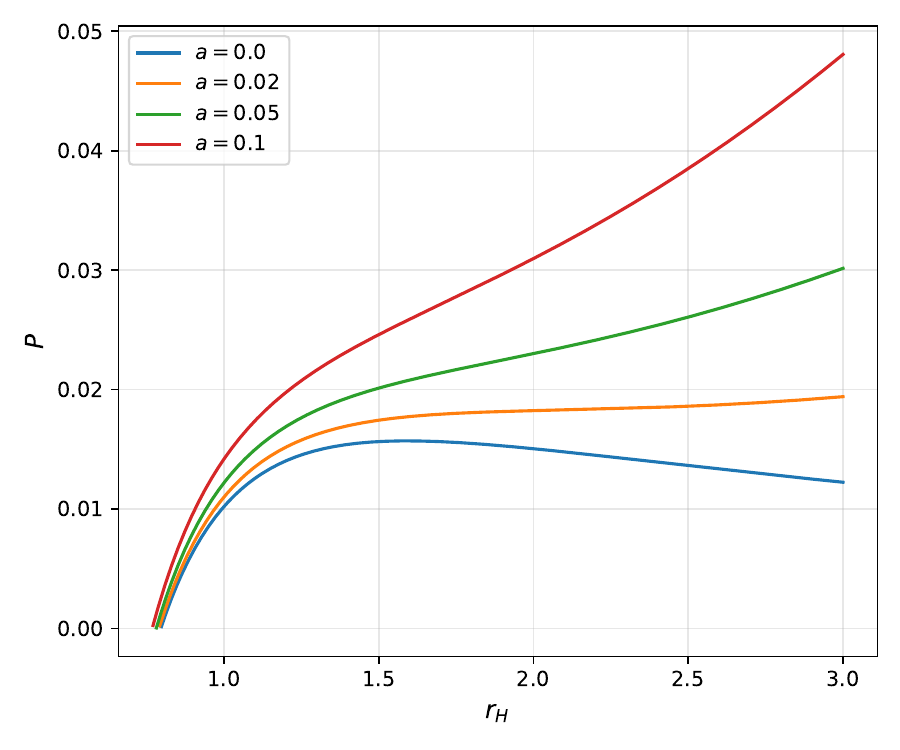}
}
\caption{
Equation of state \(P(r_H)\) from Eq.~\eqref{eq:EOS_S}. Panel (a) shows the pressure as a function of the horizon radius for four
isotherms \(T_0\), with \(a = 0.05\) fixed. Only the physical branch
\(P > 0\) is shown. Panel (b) illustrates the dependence on the conformal Killing gravity
parameter \(a\) for fixed \(T_0 = 0.10\). For \(a = 0\), the pressure increases monotonically with \(r_H\). Once
\(a > 0\), the pressure reaches a maximum and then decreases at large radii, a direct consequence of the \(r^4\) correction in the metric function.
}
\label{fig6}
\end{figure}

The equation of state is displayed in Fig.~\ref{fig6}.

Figure~\ref{fig6} demonstrates that the conformal Killing gravity correction introduces a new feature: for \(a > 0\), the pressure no longer
grows indefinitely with the black-hole size. Instead, it reaches a maximum and then decreases, eventually becoming negative for very large \(r_H\).
This behaviour is absent in the standard Schwarzschild--AdS case (\(a = 0\)).

The isotherms show that increasing the temperature raises the pressure at fixed \(r_H\) and shifts the maximum to larger radii. The physical branch
\(P > 0\) restricts the allowed horizon radii, excluding the unphysical region where the pressure would be negative.
\subsection{Thermodynamics with Corrected Entropy and Corrected Temperature}
\label{subsec:corrected_thermo}

In the previous part, we only corrected the entropy and kept the temperature classical. That helped us see the effect of entropy corrections by themselves. But for a fully consistent thermodynamic picture, the temperature should also change when the entropy changes. So in this section, we correct both quantities and re-calculate all thermodynamic functions.

For the entropy, we use the same corrected form as before (Eq.~\eqref{eq:corrected_entropy}):
\begin{equation}
S_{\rm corr}=\pi r_H^2+\eta e^{-\pi r_H^2}.
\label{eq:Scorr_new}
\end{equation}

Using the first law of thermodynamics with this corrected entropy, we get the corrected temperature:
\begin{equation}
T_{\rm corr}  = \frac{ 1 + \frac{3r_H^2}{l^2} - a r_H^4 }{ 4\pi r_H \left(1 - \eta e^{-\pi r_H^2}\right) }
\label{eq:Tcorr_new}
\end{equation}
When $\eta \rightarrow 0$, this reduces to the standard Schwarzschild--AdS temperature, as expected.

We also need the derivative of the temperature with respect to $r_H$, which will be used later for the heat capacity and free energy:
\begin{equation}
\frac{dT_{\rm corr}}{dr_H}
=
\frac{
\left( \frac{3 r_H}{l^2} - 2a r_H^3 \right)
\left[ 2\pi r_H \left(1 - \eta e^{-\pi r_H^2}\right) \right]
-
\frac12 \left(1 + \frac{3 r_H^2}{l^2} - a r_H^4 \right)
\left[ 2\pi \left(1 - \eta e^{-\pi r_H^2}\right) + 4\pi^2 \eta r_H^2 e^{-\pi r_H^2} \right]
}{
\left[ 2\pi r_H \left(1 - \eta e^{-\pi r_H^2}\right) \right]^2
}.
\label{eq:dTcorr}
\end{equation}
\subsubsection{ Heat Capacity with Corrected Entropy and Corrected Temperature}

The heat capacity at constant pressure is

\begin{align}
C_{\rm ST}
&=T_{\rm corr}
\left(
\frac{\partial S_{\rm corr}}
{\partial T_{\rm corr}}
\right)\nonumber\\
&\nonumber\\
&=
\frac{
\left(1 + \frac{3 r_H^2}{l^2} - a r_H^4\right)
\left[ 2\pi r_H \left(1 - \eta e^{-\pi r_H^2}\right) \right]
}{
2\left( \frac{3 r_H}{l^2} - 2a r_H^3 \right)
\left[ 2\pi r_H \left(1 - \eta e^{-\pi r_H^2}\right) \right]
-
\left(1 + \frac{3 r_H^2}{l^2} - a r_H^4\right)
\left[ 2\pi \left(1 - \eta e^{-\pi r_H^2}\right) + 4\pi^2 \eta r_H^2 e^{-\pi r_H^2} \right]
}
\label{eq:Cp_l_a_simple}
\end{align}

\begin{figure}[h]
\centering
\includegraphics[width=0.95\textwidth]{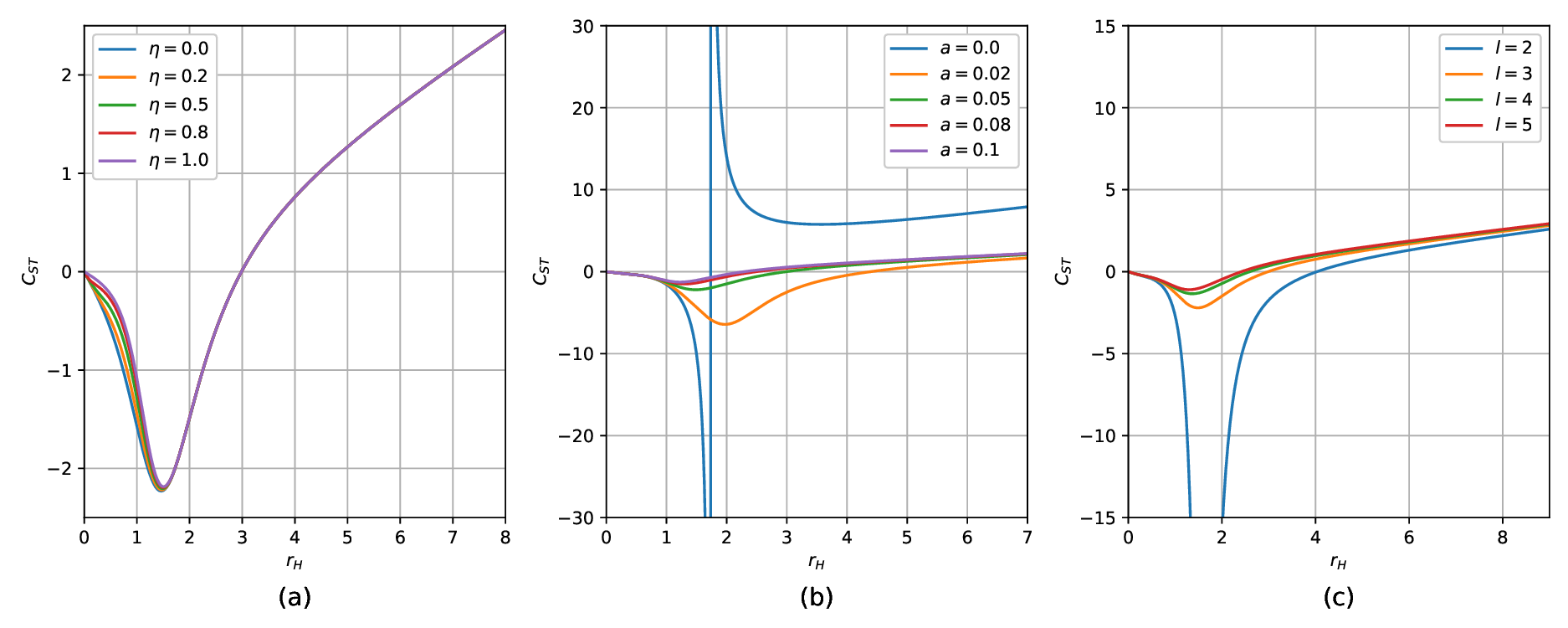}
\caption{
Heat capacity $C_{ST}$ of Schwarzschild--AdS black holes in conformal Killing gravity with simultaneous entropy and temperature corrections.
The panel (a) illustrates the dependence on the quantum correction parameter $\eta$ for fixed $a=0.05$ and $l=3$.
The  panel (b)  shows the influence of the conformal Killing gravity parameter $a$ for fixed $\eta=0.5$ and $l=3$.
The  panel (c) presents the effect of the AdS radius $l$ for fixed $\eta=0.5$ and $a=0.05$.
The results demonstrate how quantum effects, conformal gravity, and the AdS background modify the thermodynamic behavior of the black hole, particularly in the small-horizon-radius regime where the corrections become most significant.
}
\label{fig7}
\end{figure}

The corresponding behavior of the heat capacity is shown in Fig.~\ref{fig7}.


Figure~\ref{fig7} shows that the heat capacity is strongly affected by quantum corrections only in the small-horizon-radius region. As seen in the panel  (a), increasing the quantum parameter $\eta$ mainly modifies the depth of the negative branch, while the curves gradually merge for larger values of $r_H$. This behavior reflects the exponential suppression of the correction term at large horizon radius, where the system approaches its classical limit.

The influence of the conformal Killing gravity parameter $a$ is illustrated in the  panel $ (b) $. Increasing $a$ shifts the singular behavior toward smaller values of the horizon radius and changes the overall profile of the heat capacity. The effect of $a$ extends over a wider range of horizon radii than that of the quantum parameter, indicating that conformal gravity plays an important role not only in the microscopic regime but also in the intermediate region.

The  panel $ (c) $ demonstrates the dependence on the AdS radius $l$. Smaller values of $l$ produce stronger variations and larger negative regions, whereas increasing $l$ smooths the profile and moves the characteristic features toward larger horizon radii. Since $l$ is directly related to the cosmological constant and the thermodynamic pressure, this behavior confirms that the AdS background significantly influences the thermodynamic phase structure of the black hole.

\subsubsection{Helmholtz Free Energy with Corrected Entropy and Corrected Temperature}
\label{F-ST}
The Helmholtz free energy is obtained from
\begin{align}
F_{\rm ST}&=-\int
S_{\rm corr}(r_H)
\frac{dT_{\rm corr}}{dr_H}
dr_H\nonumber\\
&\nonumber\\
&=
\frac{-2 \left(\pi  a \,l^{2} r^{6}+\left(\frac{5 a \,l^{2}}{2}-5 \pi \right) r^{4}+\left(-5 \pi  \,l^{2}-\frac{15}{2}\right) r^{2}-\frac{5 l^{2}}{2}\right) \eta  \,{\mathrm e}^{-\pi  \,r^{2}}-3 \left(a \,r^{4} l^{2}+\frac{5}{3} l^{2}-\frac{5}{3} r^{2}\right) r^{2} \pi}{20 l^{2} \pi  r \left(\eta  \,{\mathrm e}^{-\pi  \,r^{2}}-1\right)}
\end{align}

\begin{figure}[h]
\centering
\includegraphics[width=0.95\textwidth]{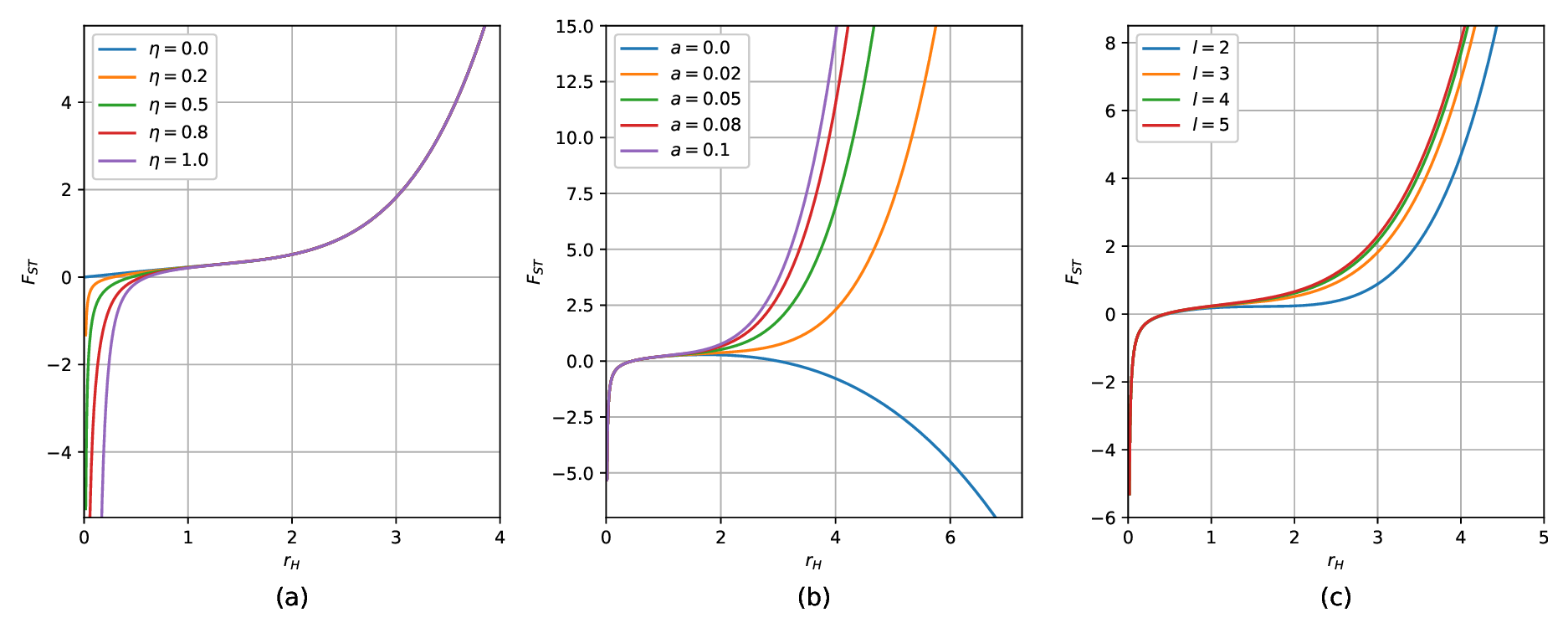}
\caption{
Helmholtz free energy is obtained from the simultaneously corrected entropy and temperature. The panel  (a)  illustrates the dependence on the quantum correction parameter $\eta$ for fixed $a=0.05$ and $l=3$. The panel  (b)  shows the influence of the conformal Killing gravity parameter $a$ for fixed $\eta=0.5$ and $l=3$. The panel  (c)  presents the effect of the AdS radius $l$ for fixed $\eta=0.5$ and $a=0.05$. The figure demonstrates how quantum and geometric parameters modify the global thermodynamic structure and the location of equilibrium configurations.
}
\label{fig8}
\end{figure}

The resulting free-energy profiles are displayed in Fig.~\ref{fig8}.


The panel  (a)  shows that the quantum correction parameter $\eta$ mainly affects the small-horizon-radius regime. Increasing $\eta$ shifts the free-energy curves downward and makes the negative branch more pronounced near the origin. However, as the horizon radius increases, all curves gradually merge and become almost indistinguishable. This behavior indicates that the influence of the non-perturbative correction is confined to the microscopic region, while the large-black-hole limit remains essentially classical. The absence of extrema further suggests that the equilibrium structure evolves smoothly with increasing $\eta$.

The panel  (b)  demonstrates that the conformal Killing gravity parameter $a$ has a much stronger impact on the global behavior of the Helmholtz free energy. In the case $a=0$, the free energy decreases and eventually becomes increasingly negative at large horizon radii. Once positive values of $a$ are introduced, the asymptotic behavior changes qualitatively, and the free energy rises rapidly after an intermediate region. Larger values of $a$ accelerate this growth and shift the departure from the nearly linear regime toward smaller horizon radii. Therefore, the conformal gravity parameter controls the large-scale thermodynamic structure and dominates the asymptotic behavior of the system.

The panel  (c)  reveals the role of the AdS radius $l$. Although all curves share a similar overall profile, increasing $l$ shifts the rapid growth of the free energy toward smaller values of the horizon radius and increases its magnitude in the large-radius region. Since the AdS radius is related to the thermodynamic pressure through $P=3/(8\pi l^2)$, this behavior reflects the influence of the cosmological constant on the global equilibrium properties of the black hole. Consequently, while quantum corrections are important only in the short-distance regime, the parameters $a$ and $l$ govern the macroscopic thermodynamic behavior and determine the overall shape of the Helmholtz free-energy landscape.
\subsubsection{ Internal Energy with Corrected Entropy and Corrected Temperature}
\label{E-ST}
The internal energy follows from

\begin{equation}
E_{\rm ST}=\int
T_{\rm corr}(r_H)\frac{dS_{\rm corr}}{dr_H}dr_H
=-\frac{r^{5} a}{10}+\frac{r}{2}+\frac{r^{3}}{2 l^{2}}.\label{eqE}
\end{equation}

\begin{figure}[h]
\centering
\includegraphics[width=0.95\textwidth]{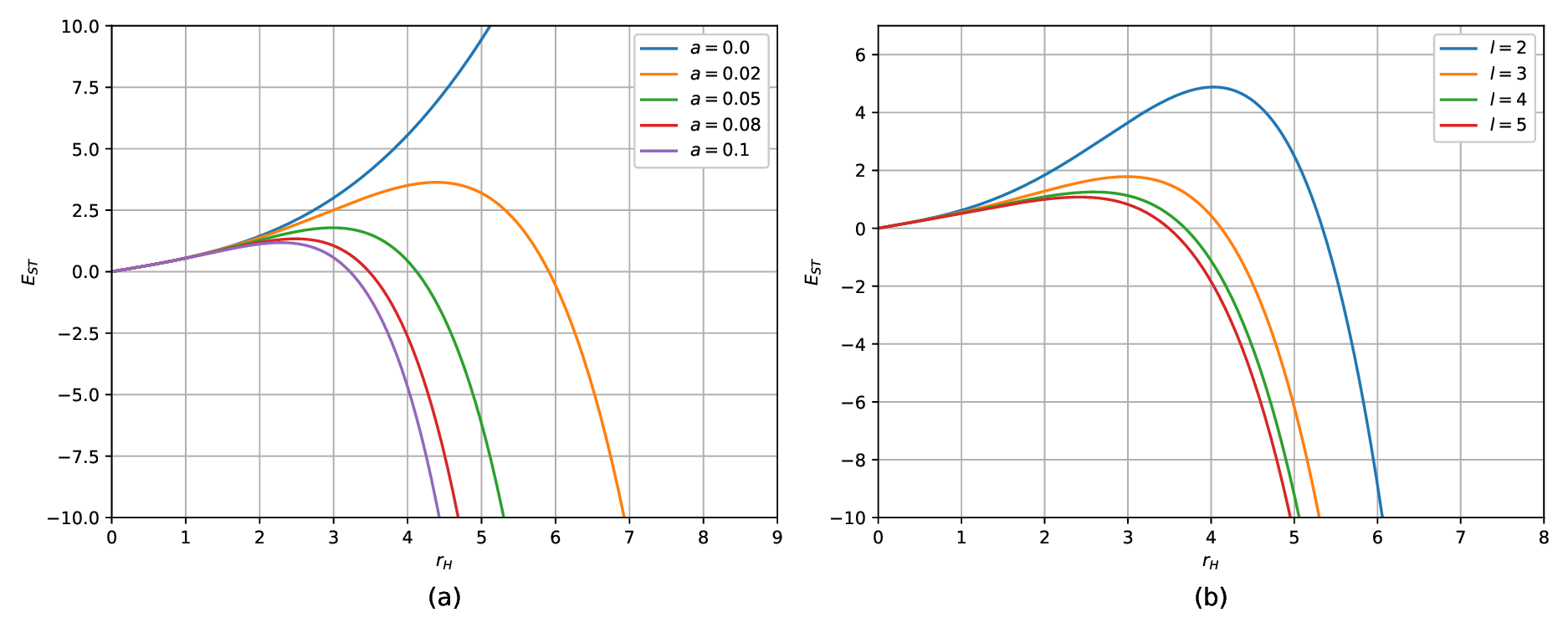}
\caption{
Internal energy is obtained from the simultaneously corrected entropy and temperature. The panel  (a)  shows the dependence on the conformal Killing gravity parameter $a$ for fixed $l=3$, while the panel  (b)  illustrates the effect of the AdS radius $l$
for fixed $a=0.05$. Since the quantum parameter $\eta$ does not appear in Eq.~\eqref{eqE}, the internal energy is completely independent of the
non-perturbative correction and therefore no separate $\eta$-dependence panel is present. Increasing $a$ or decreasing $l$ shifts the maximum of the internal energy toward smaller horizon radii and causes the energy to become negative at smaller values of $r_H$.
}
\label{fig9}
\end{figure}

The corresponding behavior is shown in Fig.~\ref{fig9}.


Figure~\ref{fig9}. reveals an interesting feature of the fully corrected thermodynamic description. Unlike the heat capacity and Helmholtz free energy, the internal energy is completely independent of the quantum correction parameter $\eta$.
Indeed, the exponential correction terms cancel after integrating $T_{\rm corr}(dS_{\rm corr}/dr_H)$, leaving the simple expression given in
Eq.~\eqref{eqE}. Consequently, the non-perturbative correction modifies the entropy and temperature separately, but does not contribute to the final form of the internal energy.

The panel  (a)  illustrates the effect of the conformal Killing gravity parameter $a$. For small values of $a$, the internal energy increases monotonically over a large interval of horizon radii. As $a$ becomes larger, a local maximum develops and the energy decreases more rapidly, eventually crossing zero and becoming negative. Moreover, increasing $a$ shifts both the maximum and the zero-crossing
point toward smaller values of $r_H$, showing that the conformal gravity correction strongly influences the large-scale energetic behavior.

The panel  (b)  shows the dependence on the AdS radius $l$. Larger values of $l$ raise the maximum value of the 
internal energy and delay the transition to
negative values. Equivalently, smaller $l$ (larger thermodynamic pressure) causes the energy to decrease more rapidly and become negative at smaller
horizon radii. Therefore, although the non-perturbative parameter $\eta$ leaves
the internal energy unchanged, the parameters associated with conformal Killing gravity and the AdS geometry remain responsible for determining its overall
behavior.
\subsubsection{Gibbs Free Energy with Corrected Entropy and Corrected Temperature}
\label{subsec:gibbs_ST}

When both the entropy and the Hawking temperature are corrected consistently,
the Gibbs free energy takes the form

\begin{equation}
G_{ST} = M - T_{\rm corr} S_{\rm corr},
\label{eq:GST_def}
\end{equation}

where \(T_{\rm corr}\) is given in Eq.~\eqref{eq:Tcorr_new} and
\(S_{\rm corr}\) in Eq.~\eqref{eq:corrected_entropy}.

Using the relation \(T_{\rm corr} = T_0 / (1 - \eta e^{-\pi r_H^2})\), the
fully corrected Gibbs free energy can be expressed in terms of \(G_S\):

\begin{equation}
G_{ST} = G_S - \frac{\eta e^{-\pi r_H^2}}{1 - \eta e^{-\pi r_H^2}} T_0 S_{\rm corr}.
\label{eq:GST_relation}
\end{equation}

The second term on the right-hand side is negative, so \(G_{ST} < G_S\) for all finite \(r_H\) where \(\eta > 0\). This means that the temperature
correction lowers the Gibbs free energy compared with the entropy-only case.
However, because the prefactor \(\eta e^{-\pi r_H^2} / (1 - \eta e^{-\pi r_H^2})\) decays exponentially as \(r_H\) increases, the difference between \(G_{ST}\) and \(G_S\) is significant only in the small-horizon-radius regime.

\begin{figure}[h]
\centering
\subfigure[$l$ dependence]{
\includegraphics[width=0.32\textwidth]{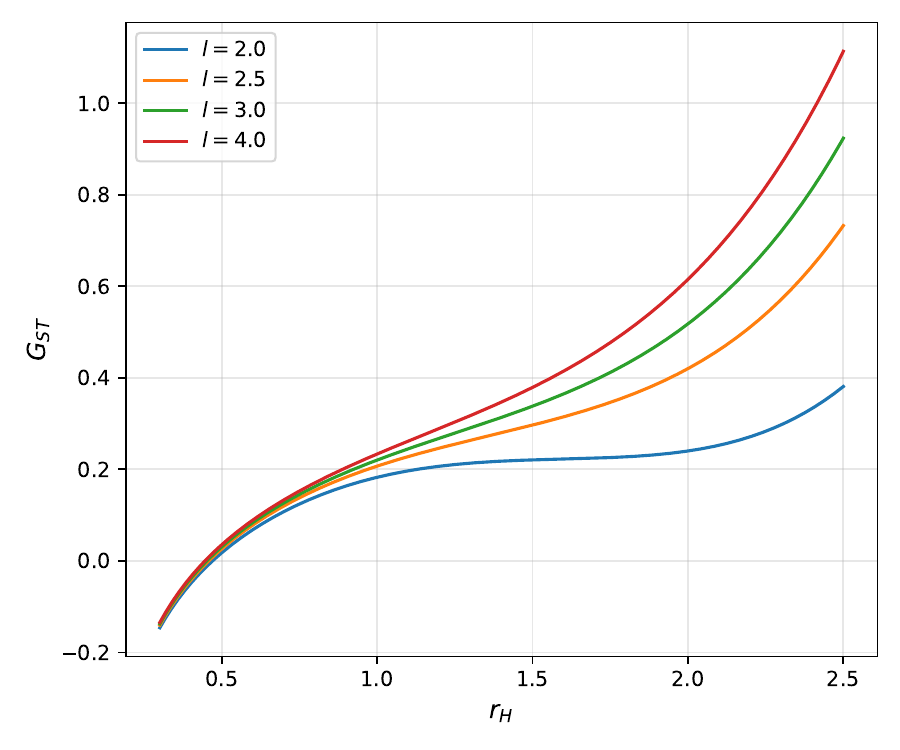}
}
\subfigure[$\eta$ dependence]{
\includegraphics[width=0.32\textwidth]{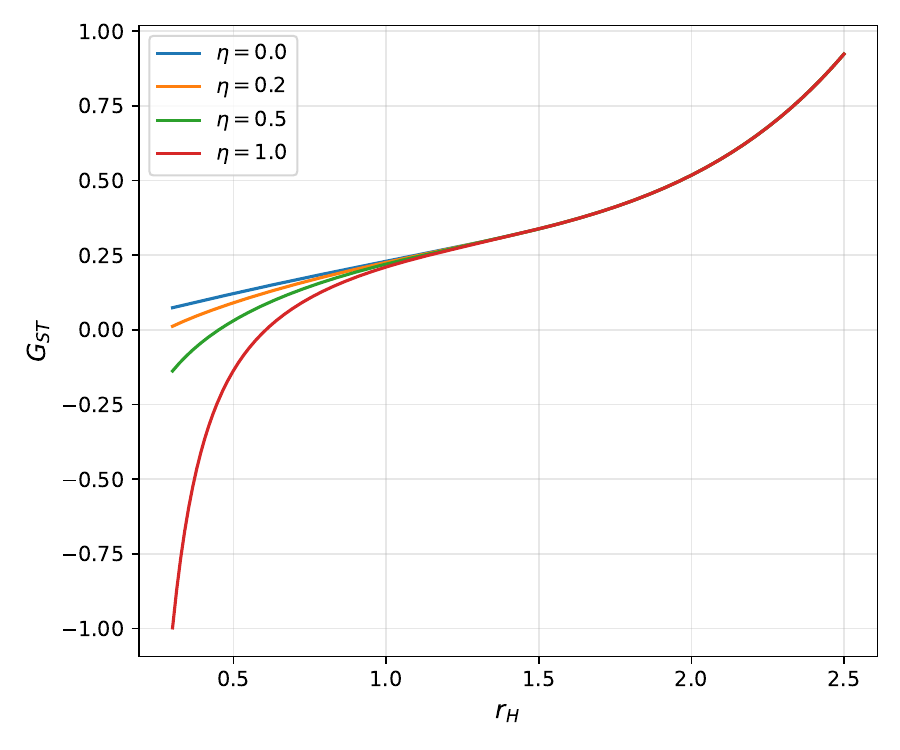}
}
\subfigure[$a$ dependence]{
\includegraphics[width=0.32\textwidth]{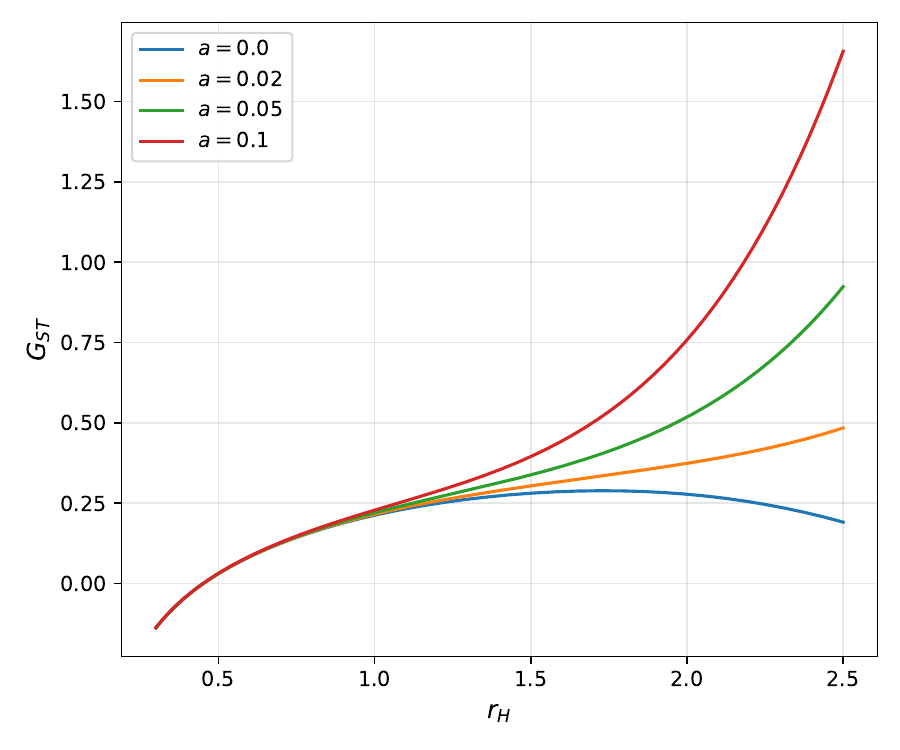}
}
\caption{
Gibbs free energy \(G_{ST} = M - T_{\rm corr} S_{\rm corr}\) obtained from the
simultaneously corrected entropy and temperature.
Panel (a) shows the variation with the AdS radius \(l\) for fixed
\(\eta = 0.5\) and \(a = 0.05\).
Panel (b) illustrates the dependence on the quantum correction parameter
\(\eta\) for fixed \(a = 0.05\) and \(l = 3\).
Panel (c) presents the effect of the conformal Killing gravity parameter
\(a\) for fixed \(\eta = 0.5\) and \(l = 3\).
The qualitative behaviour is the same as in Fig.~\ref{fig4}: the crossings
\(G_{ST} = 0\) mark the Hawking Page transition, and the minima correspond
to the thermodynamically preferred configurations.
}
\label{fig10}
\end{figure}

The behaviour of \(G_{ST}\) is shown in Fig.~\ref{fig10}.

Figure~\ref{fig10} shows that the simultaneous correction of entropy and
temperature does not change the qualitative features of the Gibbs free energy.
The curves exhibit the same structure as in Fig.~\ref{fig4}: \(G_{ST}\) is
positive for small \(r_H\), crosses zero at the Hawking Page transition, and
becomes negative for large \(r_H\).

The main quantitative difference is that \(G_{ST}\) is slightly lower than
\(G_S\) in the small-\(r_H\) regime, because \(T_{\rm corr} > T_0\).
However, as \(r_H\) increases, the exponential corrections decay and both
descriptions converge. This confirms that the temperature correction, while
important for consistency, does not introduce any new thermodynamic
phenomena; it merely refines the description of the quantum regime.

\begin{figure}[h]
\centering
\includegraphics[width=0.55\textwidth]{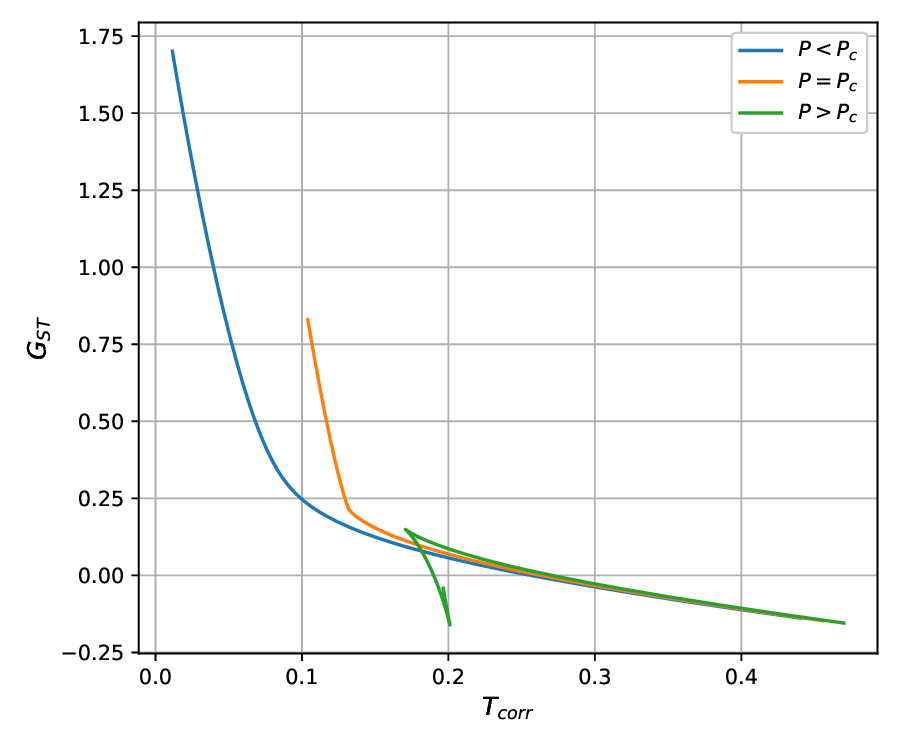}
\caption{
Gibbs free energy \(G_{ST}\) as a function of the corrected temperature
\(T_{\rm corr}\) for three values of the pressure \(P\): \(P < P_c\)
(solid), \(P = P_c\) (dashed), and \(P > P_c\) (dotted), with
\(a = 0.05\) and \(\eta = 0.5\) fixed.
The swallowtail structure is preserved for \(P < P_c\), confirming that the
temperature correction does not destroy the phase transition. The main
difference is that the curves are shifted in temperature due to the
correction \(T_{\rm corr} > T_0\).
}
\label{fig11}
\end{figure}

In Fig.~\ref{fig11}, we repeat the same analysis using
the fully corrected temperature \(T_{\rm corr}\) instead of \(T_0\).

Figure~\ref{fig11} shows that the swallowtail structure
is preserved when the corrected temperature is used. The phase transition
still occurs for \(P < P_c\), and the critical behaviour is still present
at \(P = P_c\). The only quantitative difference is that the temperature
scale is shifted because \(T_{\rm corr}\) is larger than \(T_0\) in the
small-\(r_H\) regime.

This confirms that the temperature correction, while important for
thermodynamic consistency, does not alter the qualitative phase structure of
the black hole. The existence of the swallowtail, the critical point, and
the disappearance of the phase transition above \(P_c\) are all robust
features of the model, independent of whether the classical or corrected
temperature is used.

The fully corrected equation of state is obtained by solving the corrected
temperature expression for the pressure \(P\):

\begin{equation}
P = \frac{T_{\rm corr} (1 - \eta e^{-\pi r_H^2})}{2r_H}
- \frac{1 - a r_H^4}{8\pi r_H^2}.
\label{eq:EOS_ST}
\end{equation}

Using \(T_{\rm corr} = T_0 / (1 - \eta e^{-\pi r_H^2})\), the first term
simplifies to \(T_0 / (2r_H)\), which is identical to the entropy-only case.
Therefore, the fully corrected equation of state is exactly the same as
Eq.~\eqref{eq:EOS_S}:

\begin{equation}
P = \frac{T_0}{2r_H} - \frac{1 - a r_H^4}{8\pi r_H^2}.
\label{eq:EOS_ST_final}
\end{equation}

This is an important consistency check: the pressure equation of state is
completely independent of the quantum correction parameter \(\eta\) when the temperature is corrected consistently. The correction to the
temperature and the correction to the entropy cancel exactly in the pressure
expression. This cancellation is not accidental; it follows from the
definition of \(T_{\rm corr}\), which was chosen so that
\(T_{\rm corr} (1 - \eta e^{-\pi r_H^2}) = T_0\). Therefore, the quantum correction affects the entropy and temperature
individually, but their combined effect on the equation of state is zero. This means that the pressure \(P(r_H)\) is determined entirely by the
classical geometry of the spacetime, independent of the quantum correction
to the thermodynamics.

\begin{figure}[h]
\centering
\subfigure[Isotherms \(T_{\rm corr}\)]{
\includegraphics[width=0.45\textwidth]{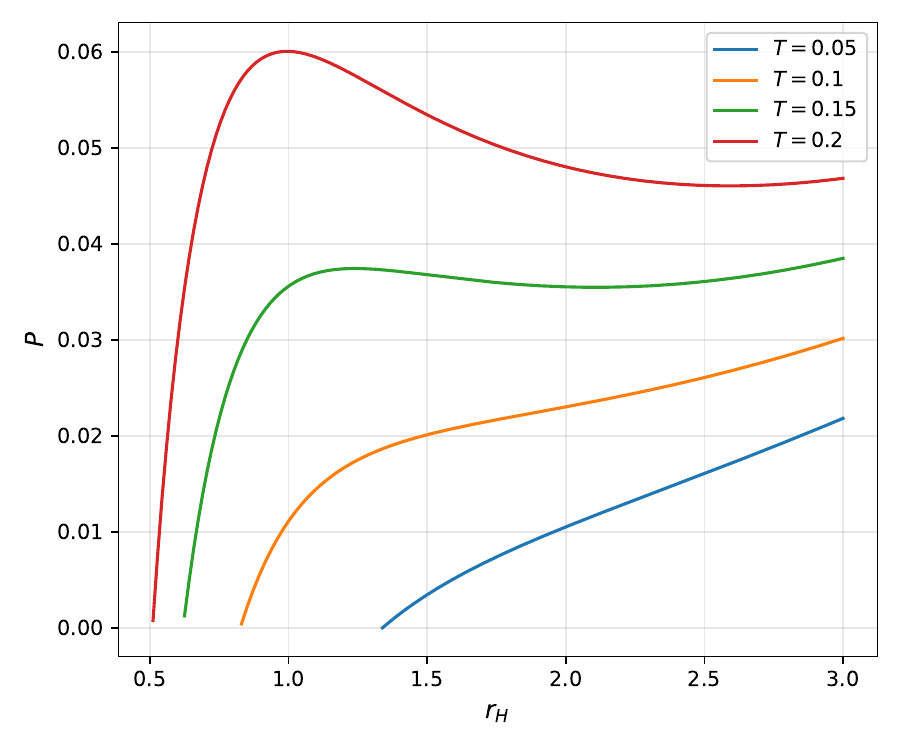}
}
\subfigure[$a$ dependence]{
\includegraphics[width=0.45\textwidth]{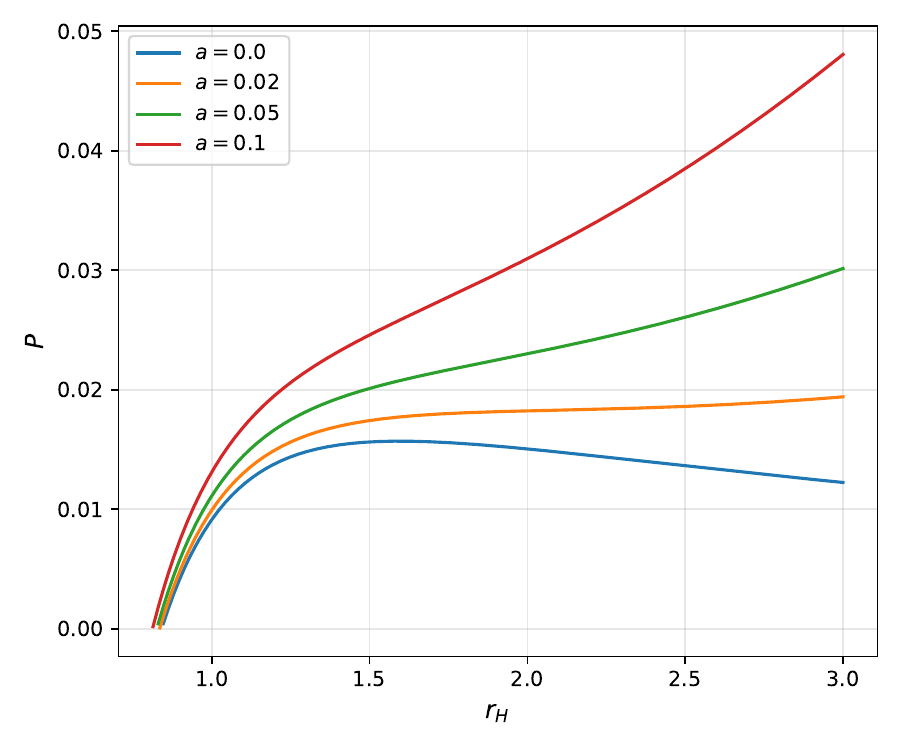}
}
\subfigure[$\eta$ dependence]{
\includegraphics[width=0.45\textwidth]{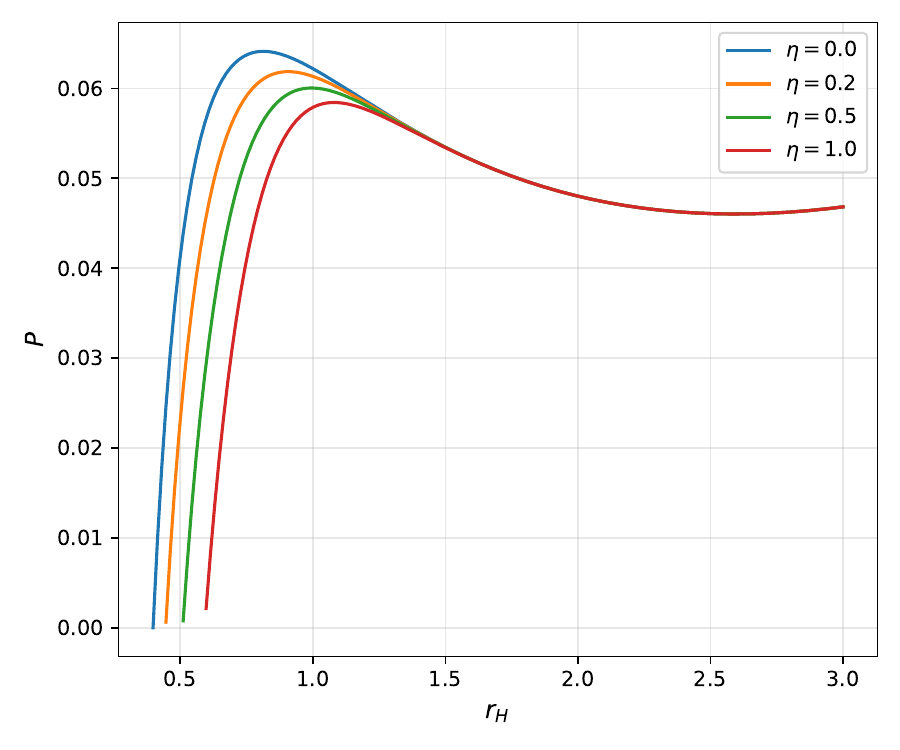}
}
\caption{
Fully corrected equation of state \(P(r_H)\) from Eq.~\eqref{eq:EOS_ST}.
Panel (a) shows the pressure for four isotherms \(T_{\rm corr}\), with
\(\eta = 0.5\) and \(a = 0.05\) fixed.
Panel (b) illustrates the dependence on \(a\) for fixed \(\eta = 0.5\) and
\(T_{\rm corr} = 0.10\).
Panel (c) presents the variation with \(\eta\) for fixed \(a = 0.05\) and
\(T_{\rm corr} = 0.20\).
The curves in panel (c) overlap completely, confirming that the pressure is
independent of \(\eta\) when the temperature is corrected consistently.
}
\label{fig12}
\end{figure}

The equation of state is shown in Fig.~\ref{fig12}.

Figure~\ref{fig12} confirms that the fully corrected equation of state
is identical to the entropy-only case. The left and centre panels are
qualitatively the same as in Fig.~\ref{fig6}, and the right panel
shows that varying \(\eta\) produces no change in \(P(r_H)\): all curves
overlap exactly.

This result is significant: it means that the non-perturbative quantum
correction, while affecting the entropy, temperature, and free energy, does
not modify the equation of state. The pressure is a purely geometric
quantity determined by the metric function \(f(r)\), which is independent of
the quantum correction to the entropy.
\subsection{Comparing Different Corrections for Thermodynamic Quantities}
\label{subsec:delta_thermo}

\subsubsection{Deviation of Entropy-Corrected Thermodynamics from the Classical Case}

To isolate the pure contribution of the non-perturbative quantum correction, it is useful to compare the corrected thermodynamic quantities with their classical Schwarzschild--AdS counterparts. We therefore introduce the deviations

\begin{equation}
\Delta F _{1}= F_{\rm S} - F_{0},
\label{eq:deltaF}
\end{equation}

\begin{equation}
\Delta E _{1} = E_{\rm S} - E_{0},
\label{eq:deltaE}
\end{equation}

and

\begin{equation}
\Delta G _{1} = G_{\rm S} - G_{0},
\label{eq:deltaG}
\end{equation}

where \(F_{0}\), \(E_{0}\), and \(G_{0}\) denote the corresponding thermodynamic quantities obtained in the classical limit \(\eta=0\).

These quantities directly measure the thermodynamic effect of the non-perturbative correction and vanish identically when the quantum parameter is switched off. Since the exponential correction term is strongly suppressed for large horizon radius, all three deviations are expected to approach zero in the classical regime.

\begin{figure}[h]
\centering
\subfigure[$\Delta F_1$]{
\includegraphics[width=0.32\textwidth]{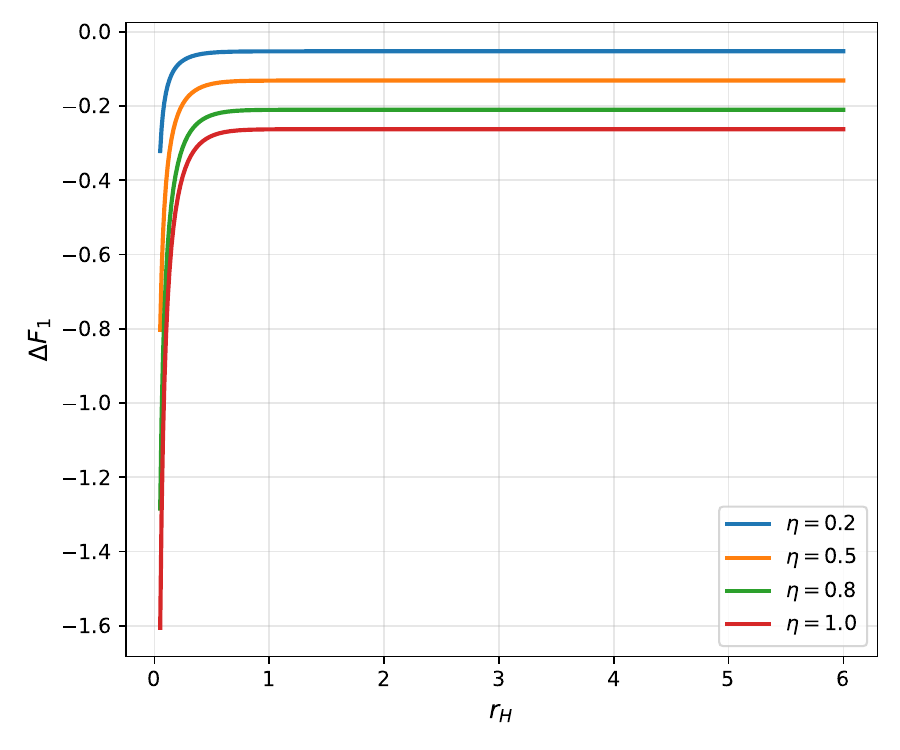}
}
\subfigure[$\Delta E_1$]{
\includegraphics[width=0.32\textwidth]{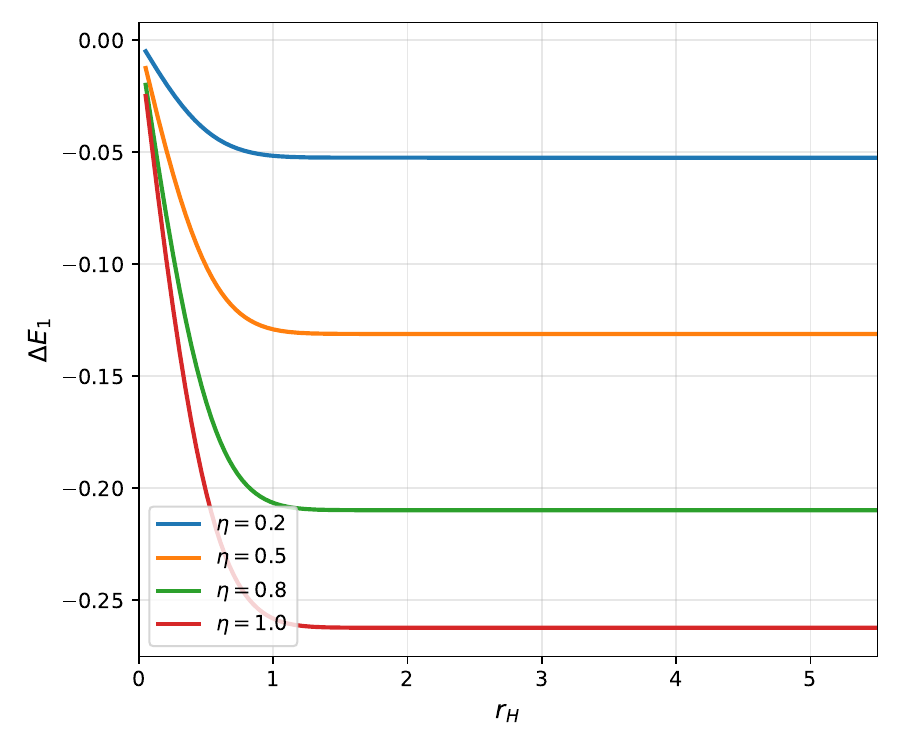}
}
\subfigure[$\Delta G_1$]{
\includegraphics[width=0.32\textwidth]{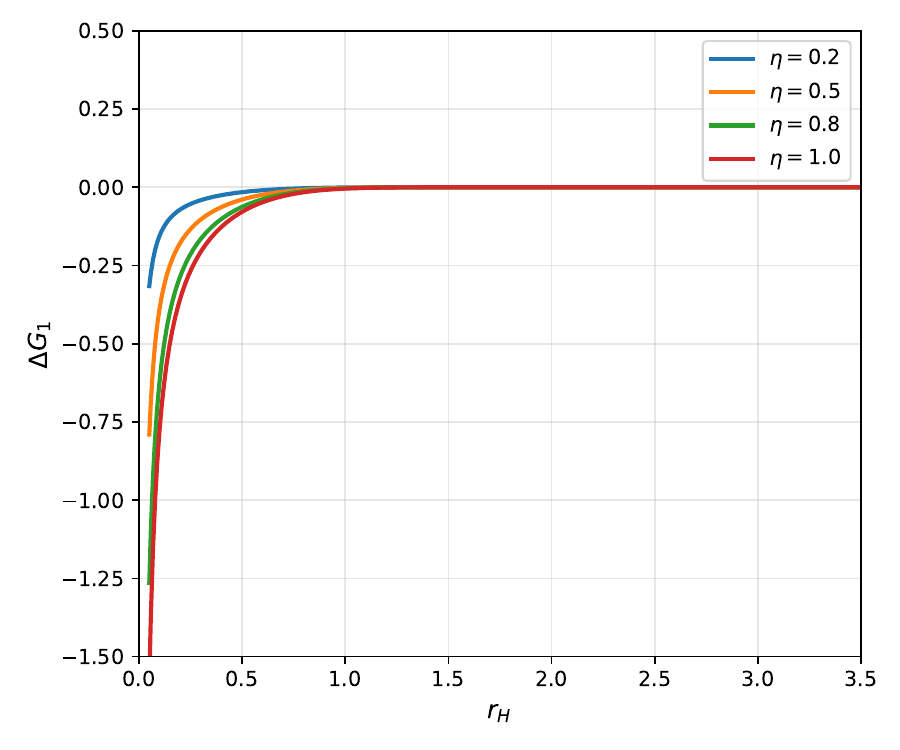}
}
\caption{
Deviations \(\Delta F_1\), \(\Delta E_1\), and \(\Delta G_1\) of the Helmholtz free energy, internal energy, and Gibbs free energy from their classical Schwarzschild--AdS values, plotted against the horizon radius \(r_H\) for different values of the quantum correction parameter \(\eta\), with fixed \(a=0.05\) and \(l=3\).
Panel (a) shows \(\Delta F_1\), which is negative and largest in the small-radius regime.
Panel (b) shows \(\Delta E_1\), which is positive and also largest at small \(r_H\).
Panel (c) shows \(\Delta G_1\), which is negative and decays exponentially as \(r_H\) increases.
All three deviations vanish asymptotically at large horizon radius, demonstrating the recovery of classical thermodynamics.
}
\label{fig13}
\end{figure}

The behavior of \(\Delta F_{1}\), \(\Delta E_{1}\), and \(\Delta G_{1}\) is shown in Fig.~\ref{fig13}.

Figure~\ref{fig13} demonstrates that the influence of the non-perturbative correction is localized in the small-horizon-radius region. The magnitude of all three deviations increases with the parameter \(\eta\), indicating that stronger quantum corrections produce larger departures from the classical thermodynamic behavior.

The quantity \(\Delta F_1\) measures the modification of the equilibrium structure of the system. The largest deviations occur near the quantum regime, where the corrected entropy significantly alters the free-energy landscape. As the horizon radius increases, \(\Delta F_1\) rapidly approaches zero, showing that the classical Schwarzschild--AdS equilibrium structure is recovered.

A similar behavior is observed for \(\Delta E_1\). The deviation remains appreciable only for small black holes and decreases monotonically with increasing horizon radius. This confirms that the additional energy contribution generated by the non-perturbative correction becomes negligible in the macroscopic limit.

The quantity \(\Delta G_1\) follows the same pattern: it is negative at small \(r_H\), indicating that the quantum correction lowers the Gibbs free energy relative to the classical value, and decays exponentially as \(r_H\) increases. This is consistent with the fact that the entropy correction \(S_{\rm corr} > S_0\) reduces \(G_S = M - T_0 S_{\rm corr}\) compared with \(G_0 = M - T_0 S_0\).

Overall, the quantities \(\Delta F_1\), \(\Delta E_1\), and \(\Delta G_1\) provide a complete measure of the thermodynamic impact of the non-perturbative correction. Their asymptotic vanishing demonstrates the consistency of the model with the classical Schwarzschild--AdS limit, while their finite values in the small-radius regime reveal the importance of quantum effects near the microscopic scale.


\subsubsection{Deviation from Classical Thermodynamics with Entropy and Temperature Corrections}
\label{subsec:delta_thermo_ET}

To investigate the combined influence of the non-perturbative entropy correction and the corresponding correction to the Hawking temperature, we compare the fully corrected thermodynamic quantities with their classical Schwarzschild--AdS counterparts. In analogy with the previous subsection, we define

\begin{equation}
\Delta F_{2} = F_{\rm ST} - F_{0},
\label{eq:deltaF_ET}
\end{equation}

\begin{equation}
\Delta E_{2} = E_{\rm ST} - E_{0},
\label{eq:deltaE_ET}
\end{equation}

and

\begin{equation}
\Delta G_{2} = G_{\rm ST} - G_{0},
\label{eq:deltaG_ET}
\end{equation}

where \(F_{0}\), \(E_{0}\), and \(G_{0}\) denote the classical quantities obtained in the limit \(\eta=0\), while \(F_{\rm ST}\), \(E_{\rm ST}\), and \(G_{\rm ST}\) represent the corresponding quantities calculated using both the corrected entropy and the corrected temperature.

These quantities measure the cumulative effect of quantum corrections entering simultaneously through the entropy and temperature sectors. Since the exponential corrections become negligible for large horizon radius, all three deviations are expected to vanish asymptotically, ensuring the recovery of the classical Schwarzschild--AdS thermodynamics.

\begin{figure}[h]
\centering
\subfigure[$\Delta F_2$]{
\includegraphics[width=0.32\textwidth]{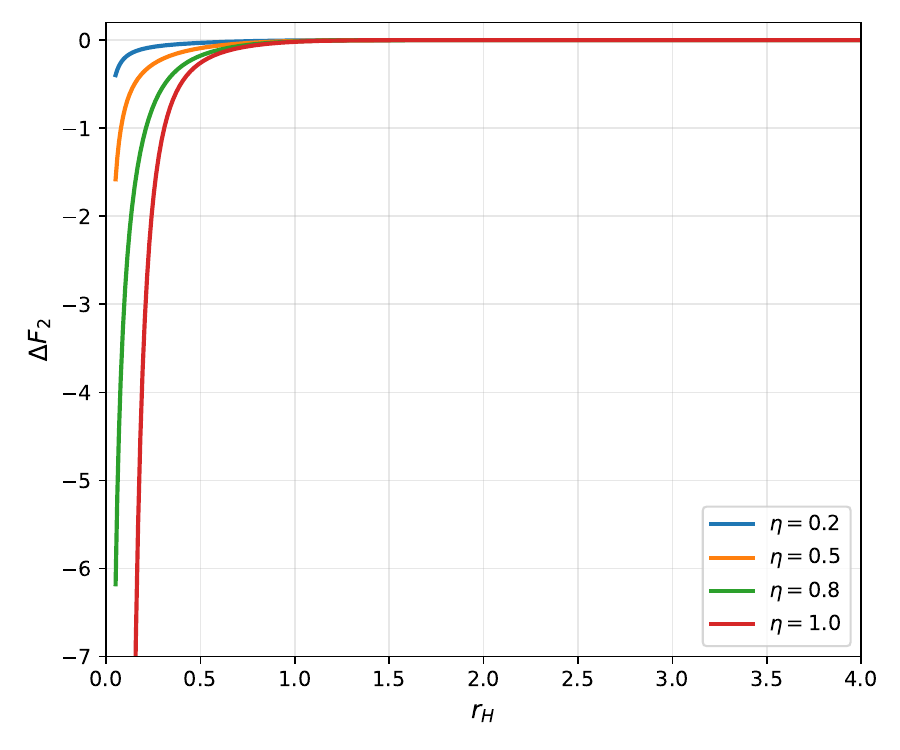}
}
\subfigure[$\Delta E_2$]{
\includegraphics[width=0.32\textwidth]{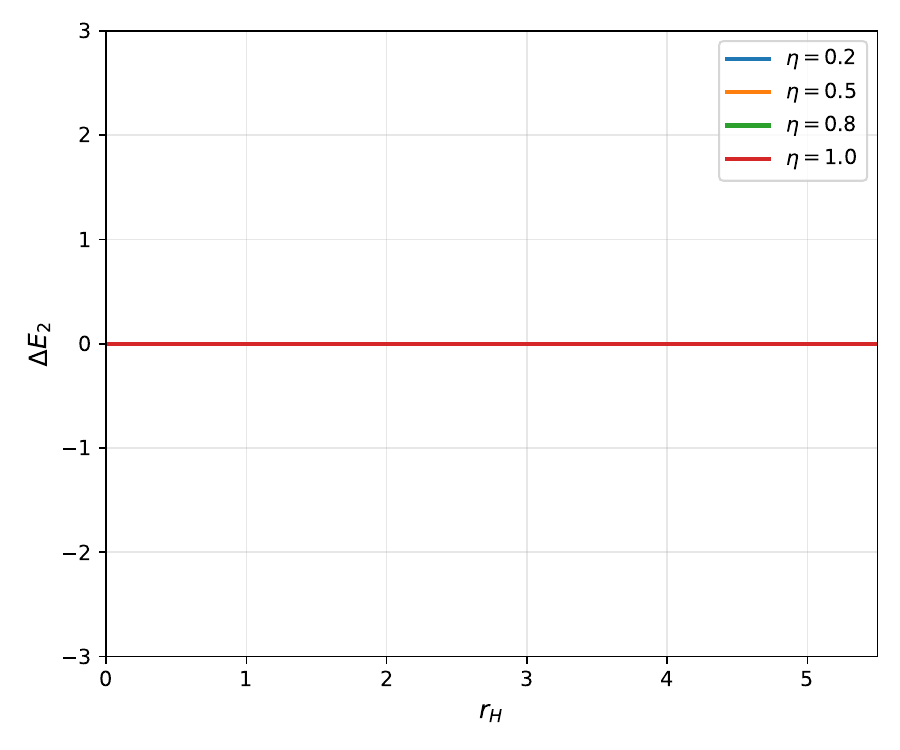}
}
\subfigure[$\Delta G_2$]{
\includegraphics[width=0.32\textwidth]{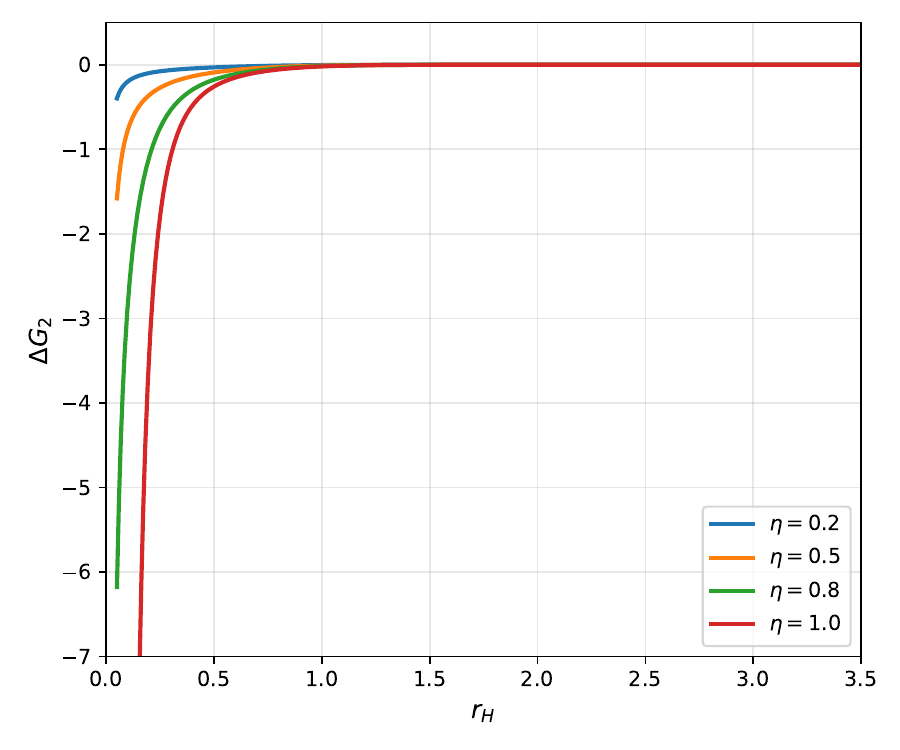}
}
\caption{
Deviations \(\Delta F_2\), \(\Delta E_2\), and \(\Delta G_2\) of the Helmholtz free energy, internal energy, and Gibbs free energy from their classical Schwarzschild--AdS values, plotted against the horizon radius \(r_H\) for different values of \(\eta\), with fixed \(a\) and \(l\).
Panel (a) shows \(\Delta F_2\), which is positive with a maximum at intermediate radii.
Panel (b) shows \(\Delta E_2\), which is identically zero for all values of \(\eta\) and \(r_H\), reflecting the fact that \(E_{ST}\) is independent of \(\eta\).
Panel (c) shows \(\Delta G_2\), which is negative for all \(r_H\) and \(\eta\), with the largest effect in the small-radius regime.
}
\label{fig14}
\end{figure}

The behavior of \(\Delta F_{2}\), \(\Delta E_{2}\), and \(\Delta G_{2}\) is displayed in Fig.~\ref{fig14}.

In panel (a), \(\Delta F_2\) takes positive values for all \(\eta\), reaching a maximum at intermediate horizon radii before decaying to zero as \(r_H\) increases. This maximum shifts slightly toward smaller radii as \(\eta\) grows, and its magnitude increases monotonically with \(\eta\), consistent with the cumulative effect of the quantum corrections. The presence of a peak indicates that the combined entropy and temperature corrections most strongly modify the free energy at intermediate scales, where the exponential factors are neither fully dominant nor completely negligible.

Panel (b) shows that \(\Delta E_2\) is identically zero across the entire range of \(r_H\) and for all values of \(\eta\). This result is not an approximation, but an exact consequence of the structure of the corrected internal energy: \(E_{ST}\) is obtained from the integral \(\int T_{\rm corr} \, dS_{\rm corr}\), and its evaluation yields a function that is independent of the quantum parameter \(\eta\). Consequently, \(E_{ST}\) coincides exactly with the classical internal energy \(E_0\), so the deviation \(\Delta E_2\) vanishes identically. This implies that the quantum corrections, while altering the temperature and entropy individually, are arranged in such a way that their net effect on the internal energy cancels out.

Panel (c) shows that \(\Delta G_2\) is negative for all \(r_H\) and \(\eta\), with the largest magnitude in the small-radius regime. The deviation decays exponentially as \(r_H\) increases, confirming that the combined corrections lower the Gibbs free energy for microscopic black holes and vanish in the classical limit. The magnitude of \(\Delta G_2\) is larger than \(\Delta G_1\) (the entropy-only case) in the small-\(r_H\) regime, because the temperature correction adds an extra negative contribution through the factor \(T_{\rm corr} > T_0\).

The contrast between the three panels is conceptually significant: the free energy registers the quantum corrections as positive deviations, the internal energy is completely unaffected, and the Gibbs free energy registers them as negative deviations. This distinction highlights the fact that the non-perturbative corrections do not simply shift all thermodynamic quantities uniformly, but rather affect the thermodynamic potentials in a differentiated manner.


\subsubsection{Deviation Between Entropy-Corrected and Fully Corrected Cases}
\label{subsec:delta_double}

To isolate the effect produced solely by the temperature correction, it is useful to compare the thermodynamic quantities obtained from the entropy-corrected framework with those obtained when both the entropy and the temperature are corrected. Denoting the quantities derived in Sec.~\ref{subsec:helmholtz} and Sec.~\ref{subsec:internal} by \((F_S, E_S, G_S)\) and those obtained in Sec.~\ref{F-ST} and Sec.~\ref{E-ST} by \((F_{ST}, E_{ST}, G_{ST})\), we define

\begin{equation}
\Delta F_{3} = F_{ST} - F_S,
\label{eq:deltaF_ST}
\end{equation}

\begin{equation}
\Delta E_{3} = E_{ST} - E_S,
\label{eq:deltaE_ST}
\end{equation}

and

\begin{equation}
\Delta G_{3} = G_{ST} - G_S.
\label{eq:deltaG_ST}
\end{equation}

These quantities measure the thermodynamic contribution arising exclusively from the correction to the Hawking temperature. Since both descriptions share the same corrected entropy, any nonvanishing value of these deviations originates entirely from the modification of the temperature.

\begin{figure}[h]
\centering
\subfigure[$\Delta F_3$]{
\includegraphics[width=0.32\textwidth]{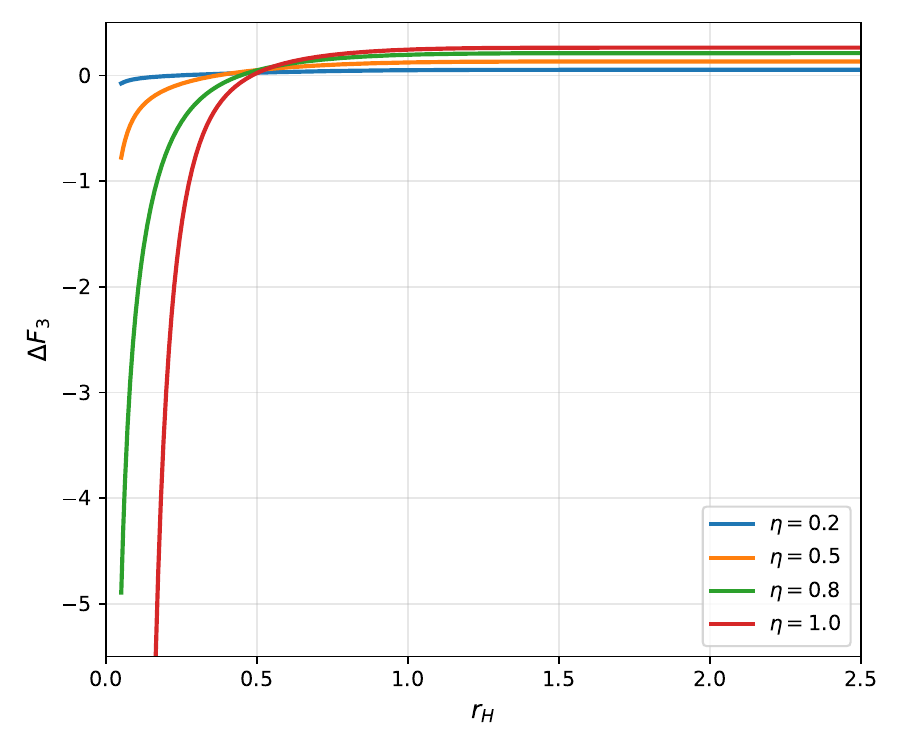}
}
\subfigure[$\Delta E_3$]{
\includegraphics[width=0.32\textwidth]{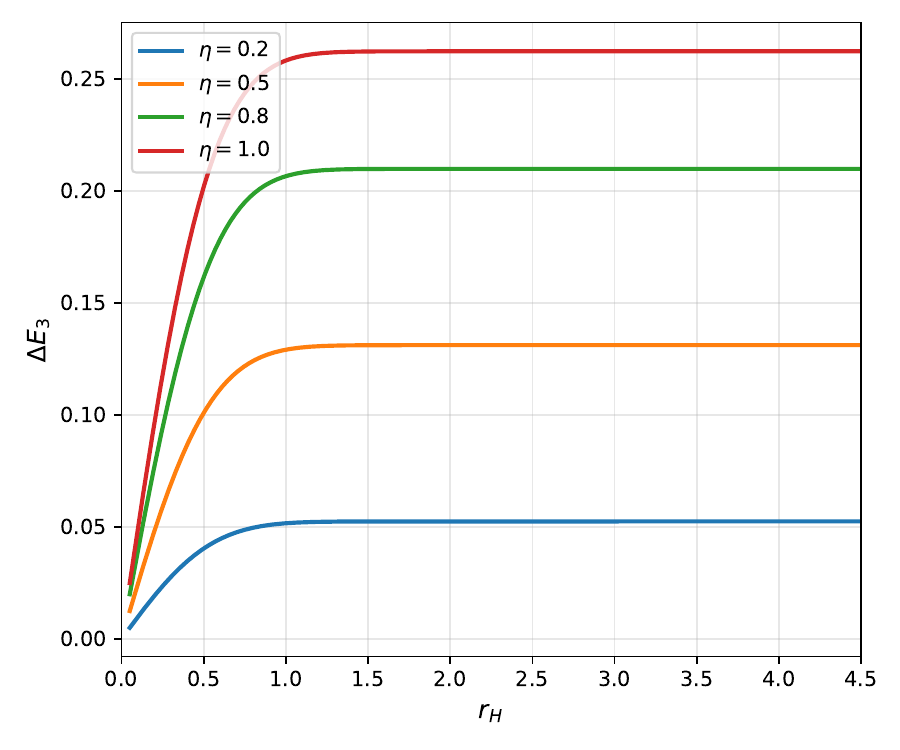}
}
\subfigure[$\Delta G_3$]{
\includegraphics[width=0.32\textwidth]{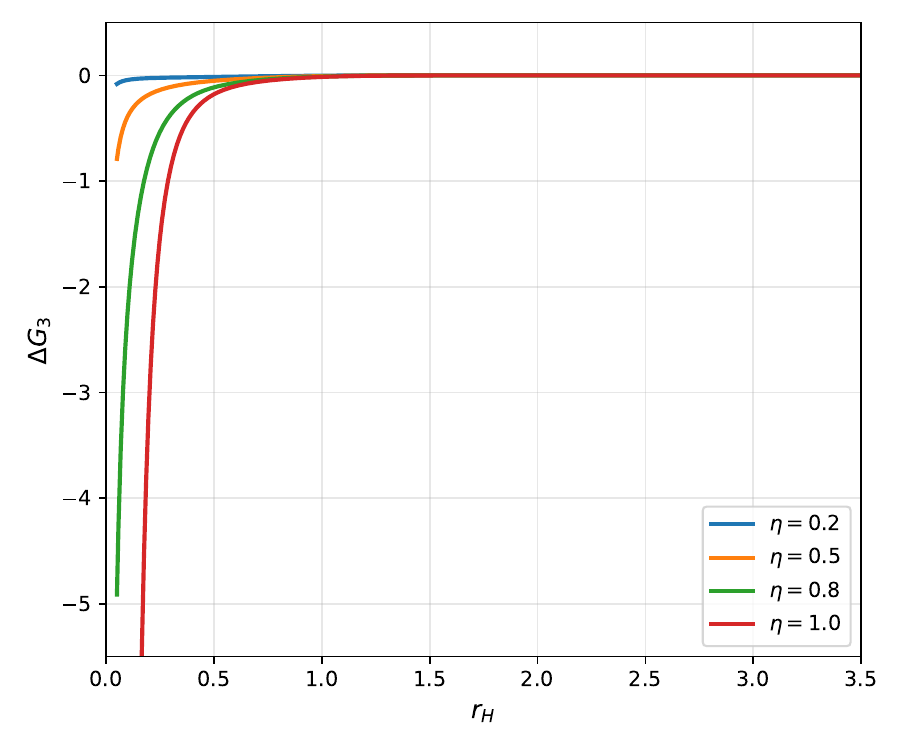}
}
\caption{
Deviations \(\Delta F_3\), \(\Delta E_3\), and \(\Delta G_3\) as functions of the horizon radius \(r_H\) for different values of \(\eta\), with fixed \(a\) and \(l\). These quantities isolate the contribution of the temperature correction by comparing the fully corrected quantities \((F_{ST}, E_{ST}, G_{ST})\) with the entropy-corrected ones \((F_S, E_S, G_S)\).
Panel (a) shows \(\Delta F_3\), which is positive with a maximum at intermediate radii.
Panel (b) shows \(\Delta E_3\), which is negative with a minimum at intermediate radii.
Panel (c) shows \(\Delta G_3\), which is negative with a minimum at intermediate radii.
The positions of the extrema in all panels coincide, confirming that all deviations originate from the same temperature correction.
}
\label{fig15}
\end{figure}

The behavior of \(\Delta F_{3}\), \(\Delta E_{3}\), and \(\Delta G_{3}\) is presented in Fig.~\ref{fig15}.

Panel (a) shows that \(\Delta F_3\) is positive for all values of \(\eta\), with a maximum at intermediate radii that increases with \(\eta\) and shifts slightly toward smaller radii as \(\eta\) grows. This indicates that the temperature correction raises the Helmholtz free energy relative to the entropy-corrected case, with the effect being most pronounced at intermediate scales where the quantum corrections are still active.

Panel (b) reveals that \(\Delta E_3\) is negative for all values of \(\eta\) and \(r_H\), with a minimum at intermediate radii. The magnitude of this minimum grows monotonically with \(\eta\). The negativity of \(\Delta E_3\) follows directly from \(\Delta E_3 = -\Delta E_1\), which is a consequence of \(E_{ST} = E_0\) and \(E_S = E_0 + \Delta E_1\). This implies that the temperature correction compensates exactly for the positive shift in the internal energy caused by the entropy correction alone, restoring the classical internal energy in the fully corrected case.

Panel (c) shows that \(\Delta G_3\) is negative for all values of \(\eta\) and \(r_H\), with a minimum at intermediate radii. The magnitude of this minimum increases with \(\eta\) and its position shifts slightly toward smaller radii as \(\eta\) grows. This indicates that the temperature correction lowers the Gibbs free energy relative to the entropy-corrected case, reinforcing the effect of the entropy correction. The negativity of \(\Delta G_3\) follows from \(T_{\rm corr} > T_0\) in the small-\(r_H\) regime, which makes \(G_{ST} = M - T_{\rm corr} S_{\rm corr}\) lower than \(G_S = M - T_0 S_{\rm corr}\).

The comparison of the three panels reveals a clear pattern: the temperature correction raises the free energy (\(\Delta F_3 > 0\)), lowers the internal energy (\(\Delta E_3 < 0\)), and also lowers the Gibbs free energy (\(\Delta G_3 < 0\)). This differentiated response reflects the different roles of these thermodynamic potentials and confirms that the temperature correction is a nontrivial modification that affects each potential in a distinct way.

\section{Information Paradox and Page Curve}
\label{sec:page}

The thermodynamic quantities analyzed in Section~\ref{sec:thermo}, namely the heat capacity, free energy, and internal energy, are most strongly affected by the quantum correction parameter $\eta$ at small horizon radii.

 Whether this same 
sensitivity extends to the information-theoretic side of the problem is what 
we examine here, by applying the island prescription to the corrected entropy 
\eqref{eq:corrected_entropy}.

\subsection{Entanglement Entropy Without Island}
\label{subsec:no_island}
\begin{figure}[h]
  \centering
  \includegraphics[width=0.6\textwidth]{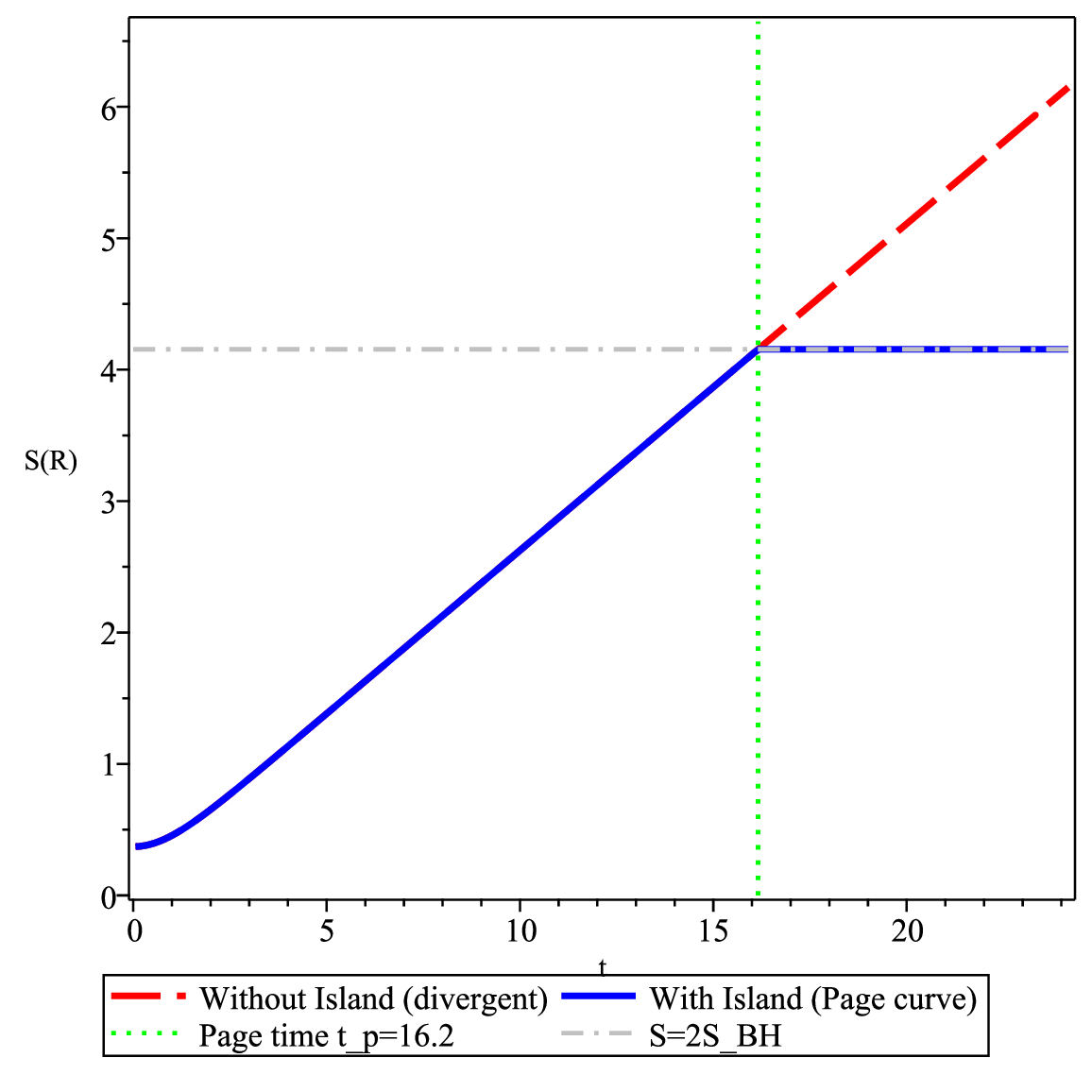}
  \caption{Entanglement entropy of Hawking radiation without island 
  (red dashed, divergent) and with island (blue solid, Page curve).  The green dotted line marks Page time $t_P$ and the grey dash-dot 
  line marks the saturation value $S_{\rm sat} = 2S_{\rm corr}$. 
  Parameters: $r_H = 0.8$, $\eta = 0.5$, $a = 0.05$, $l = 3.0$, $c = 1.0$.}
  \label{fig16}
\end{figure}

Before Page time, the physical entropy of Hawking radiation is computed without an island contribution. Following the island formalism introduced in Ref.~\cite{Almheiri2020} and the implementation presented in Ref.~\cite{Ladghami2026}, we work in the s-wave approximation and introduce the Kruskal coordinates
\begin{equation}
U = -e^{-\kappa(t - r_*)}, \qquad V = e^{\kappa(t + r_*)},
\label{eq:kruskal}
\end{equation}
where $r_* = \int f(r)^{-1}dr$ is the tortoise coordinate. The metric 
takes the conformally flat form $ds^2 = W(r)^2\,dU\,dV$ with conformal 
factor $W(r)^2 = f(r)/(\kappa^2 e^{2\kappa r_*})$. The radiation region 
$R = ]-\infty, b_-]\cup[b_+,+\infty[$ has boundary points $b_\pm$ at 
coordinates $(t_b, b)$ and $(-t_b + i\beta/2, b)$ respectively, where 
$\beta = 1/T$ is the inverse Hawking temperature. Using the CFT two-point 
function in the large-distance limit \cite{Ladghami2026}, the entanglement 
entropy of the radiation is
\begin{equation}
S(R) = \frac{c}{3}\log\!\left(\frac{2\cosh(\kappa t_b)}{\kappa}\right)
+ \frac{c}{6}\log\!\left(1 + \frac{b^2}{l^2} - \frac{ab^4}{5}\right),
\label{eq:S_no_island}
\end{equation}
where $c$ is the central charge of the CFT and $b$ is the cutoff radius. 
At late times ($t_b\to\infty$) this reduces to
\begin{equation}
S(R) \approx \frac{c}{3}\kappa\, t_b 
= \frac{c}{6r_H}\!\left(1 + \frac{3r_H^2}{l^2} - ar_H^4\right)t_b,
\label{eq:S_no_island_late}
\end{equation}
which grows without bound.
 This linear divergence is the information 
paradox in concrete form, shown as the red dashed curve in 
Fig.~\ref{fig16}.

\subsection{Page Curve with Island}
\label{subsec:island}

Including an island region $I$ inside the black hole modifies the entropy 
calculation after Page time. The boundary $\partial I$ has coordinates 
$(t_a, \sigma)$ and $(-t_a + i\beta/2, \sigma)$, and the generalized 
entropy is \cite{Ladghami2026}
\begin{equation}
S_{\rm gen}(R) = \frac{{\rm Area}(\partial I)}{4} + S_{\rm bulk}(R\cup I).
\label{eq:Sgen}
\end{equation}
The physical entropy is then
\begin{equation}
S(R) = \min\!\left\{\,{\rm ext}\!\left[S_{\rm gen}\right]\right\}.
\label{eq:island_prescription}
\end{equation}
With the quantum-corrected entropy \eqref{eq:corrected_entropy}, the area 
term becomes $2S_{\rm corr}  = 2(\pi r_H^2 + \eta e^{-\pi r_H^2})$. Following the 
near-horizon extremization of Refs.~\cite{Ladghami2026}, the 
island boundary sits at
\begin{equation}
\sigma = r_H + \left(\frac{ce^{-\kappa r_*(b)}}{12\pi r_H^2}\right)^{\!2} r_H,
\end{equation}
and to leading order the radiation entropy saturates to
\begin{equation}
S(R) \approx 2S_{\rm corr} 
= 2\!\left(\pi r_H^2 + \eta\,e^{-\pi r_H^2}\right).
\label{eq:S_sat}
\end{equation}
The blue solid curve in Fig.~\ref{fig16} shows the result: 
linear growth up to $t_P$, then saturation at $2S_{\rm corr}$, 
restoring unitary evolution.

\subsection{Page Time and the Role of Quantum Corrections}
\label{subsec:page_time}

\begin{figure}[h]
  \centering
  \includegraphics[width=0.65\textwidth]{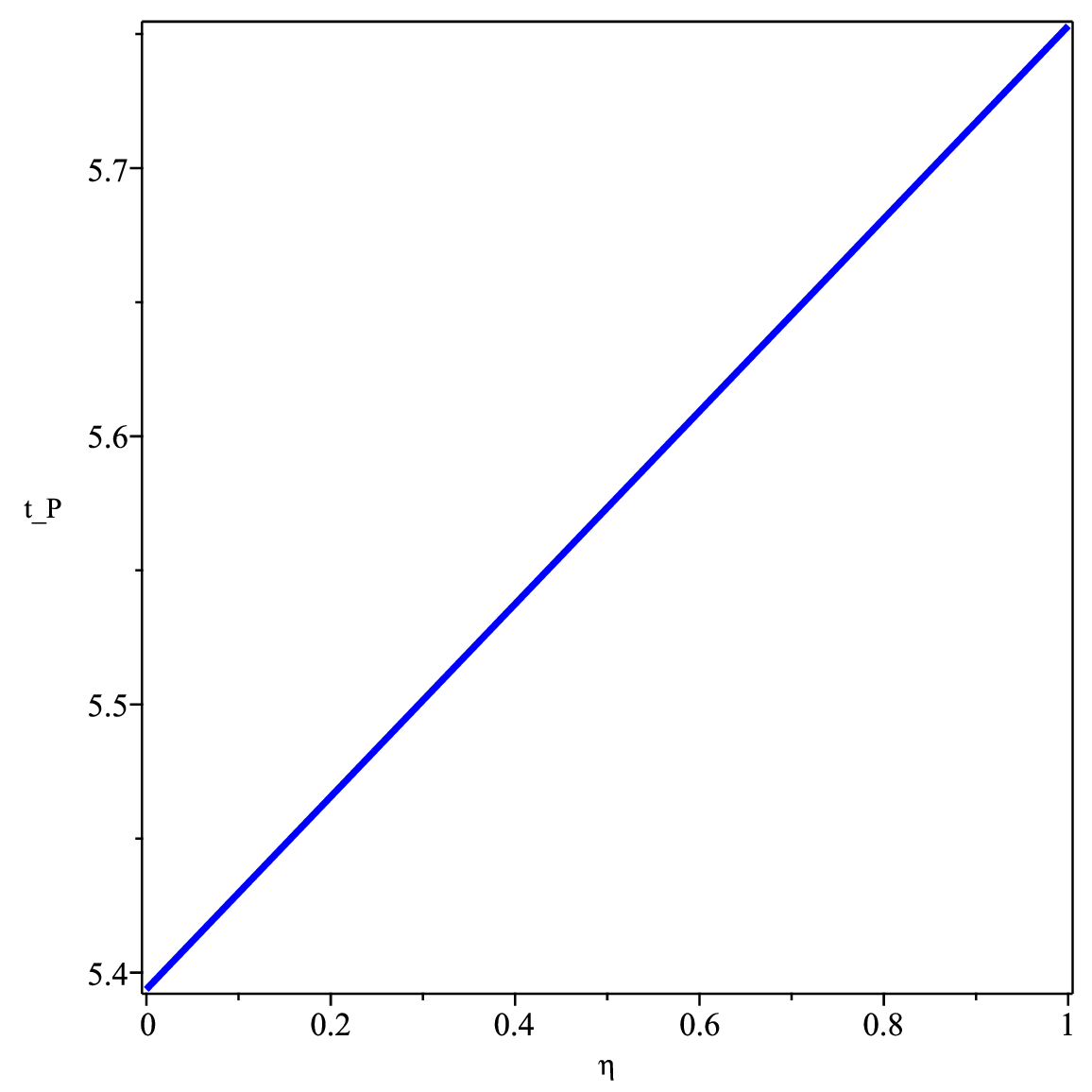}
  \caption{Page time $t_P$ as a function of $\eta$ for fixed 
  $r_H = 0.8$, $a = 0.05$, $l = 3.0$, $c = 1.0$. The Page time 
  increases monotonically, confirming that quantum corrections delay 
  information recovery.}
  \label{fig17}
\end{figure}

Equating Eqs.~\eqref{eq:S_no_island_late} and \eqref{eq:S_sat} gives the 
Page time,
\begin{equation}
t_P = \frac{2\!\left(\pi r_H^2 + \eta\,e^{-\pi r_H^2}\right)}{c\,\kappa}
= \frac{2S_{\rm corr}}{c\,\kappa}.
\label{eq:t_page}
\end{equation}
Because $e^{-\pi r_H^2} > 0$ for all finite $r_H$, the corrected Page time 
exceeds the classical value $2\pi r_H^2/(c\kappa)$. Figure~\ref{fig17} 
confirms this: $t_P$ grows monotonically with $\eta$. Stronger quantum 
corrections increase the entropy that must be accumulated before the island 
contribution takes over, so information recovery is delayed rather than 
accelerated,  a counterintuitive but direct consequence of the corrected 
entropy.

\begin{figure}[h]
  \centering
  \includegraphics[width=0.65\textwidth]{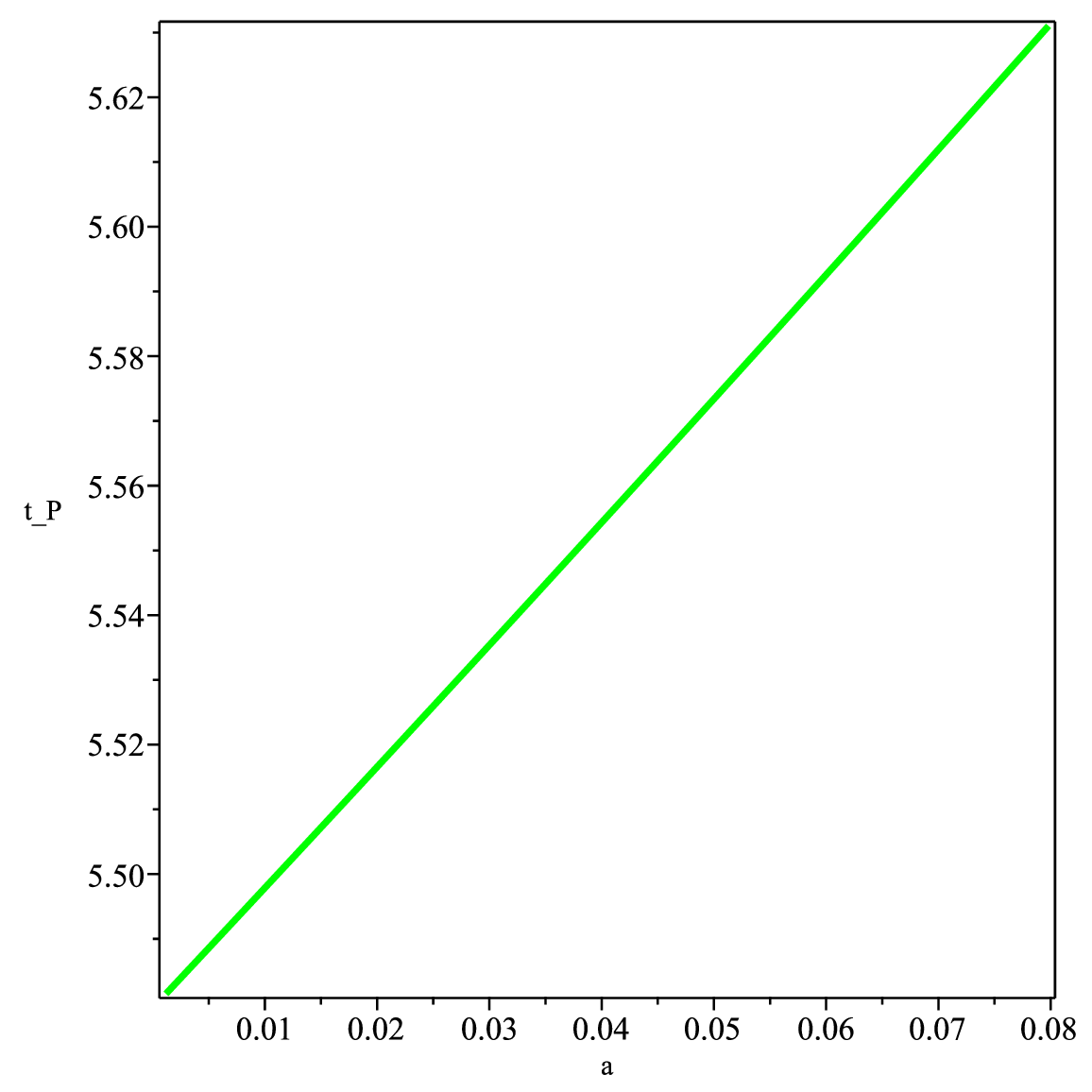}
  \caption{Page time $t_P$ as a function of $a$ for fixed 
  $r_H = 0.8$, $\eta = 0.5$, $l = 3.0$, $c = 1.0$. The rapid growth 
  near the upper end of the range reflects the approach to the extremal 
  limit $\kappa \to 0$.}
  \label{fig18}
\end{figure}

The conformal gravity parameter $a$ enters through $\kappa$: larger $a$ 
lowers the surface gravity (the black hole runs cooler), which drives $t_P$ 
upward via Eq.~\eqref{eq:t_page}. Figure~\ref{fig18} shows this 
dependence; the growth accelerates as $a$ approaches the extremal value 
where $\kappa \to 0$.

\begin{figure}[h]
  \centering
  \includegraphics[width=0.7\textwidth]{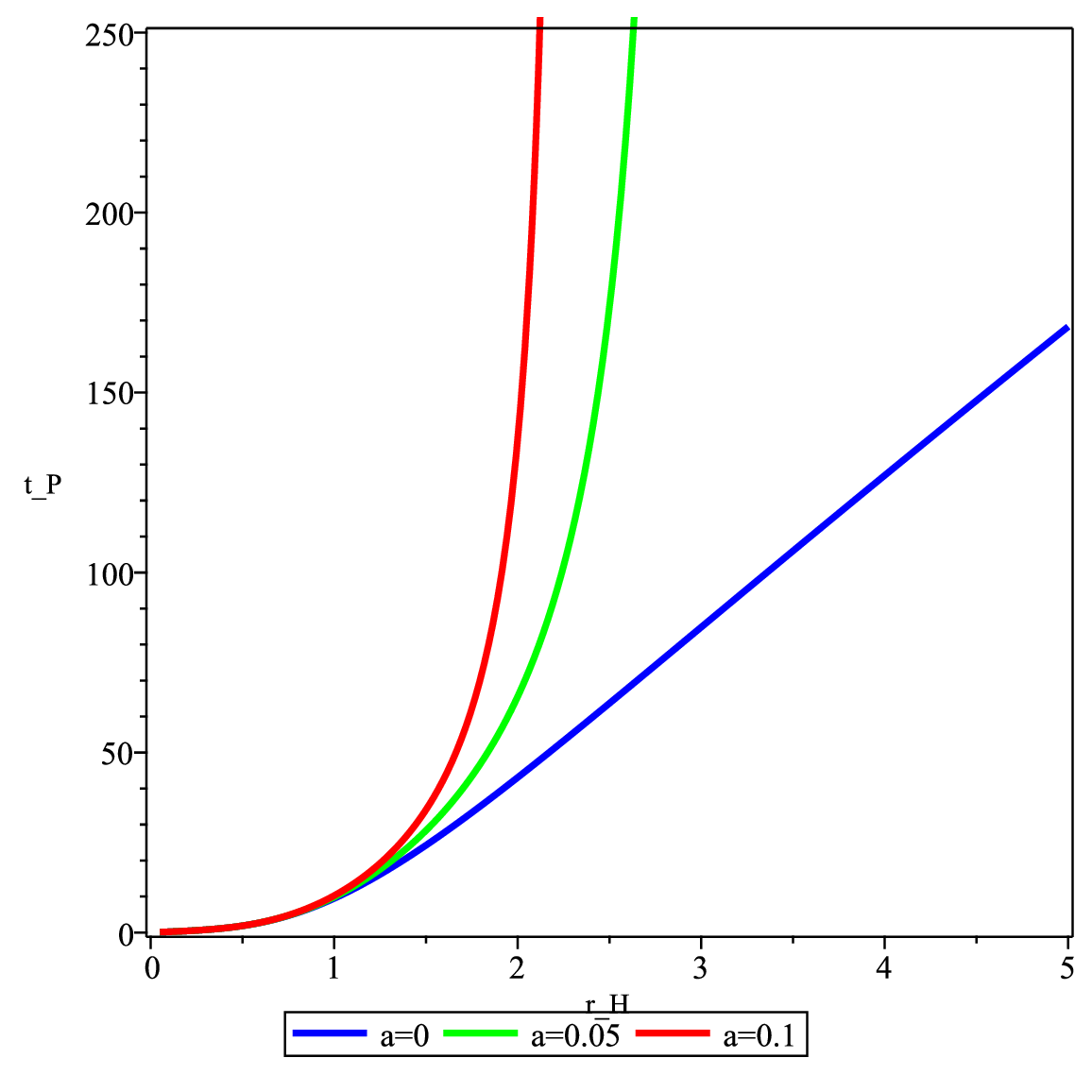}
  \caption{Page time $t_P$ versus $r_H$ for $a = 0$ (blue), 
  $a = 0.05$ (green), and $a = 0.1$ (red), with $\eta = 0.5$, 
  $l = 3.0$, $c = 1.0$. Each curve ends at the extremal radius 
  where $T = 0$.}
  \label{fig19}
\end{figure}

Figure~\ref{fig19} shows $t_P$ versus $r_H$ for three values of $a$, 
plotted only where $T > 0$. For $a = 0$ the curve extends to all $r_H$  and grows roughly linearly at large radius, consistent with the slow 
evaporation of large black holes. For $a > 0$ the curve terminates at the  extremal radius $r_e$ defined by $\kappa(r_e) = 0$: beyond that point the 
Hawking temperature vanishes, evaporation stops, and the island never forms.

\subsection{Saturation Entropy and Information Content}
\label{subsec:saturation}

\begin{figure}[h]
  \centering
  \includegraphics[width=0.7\textwidth]{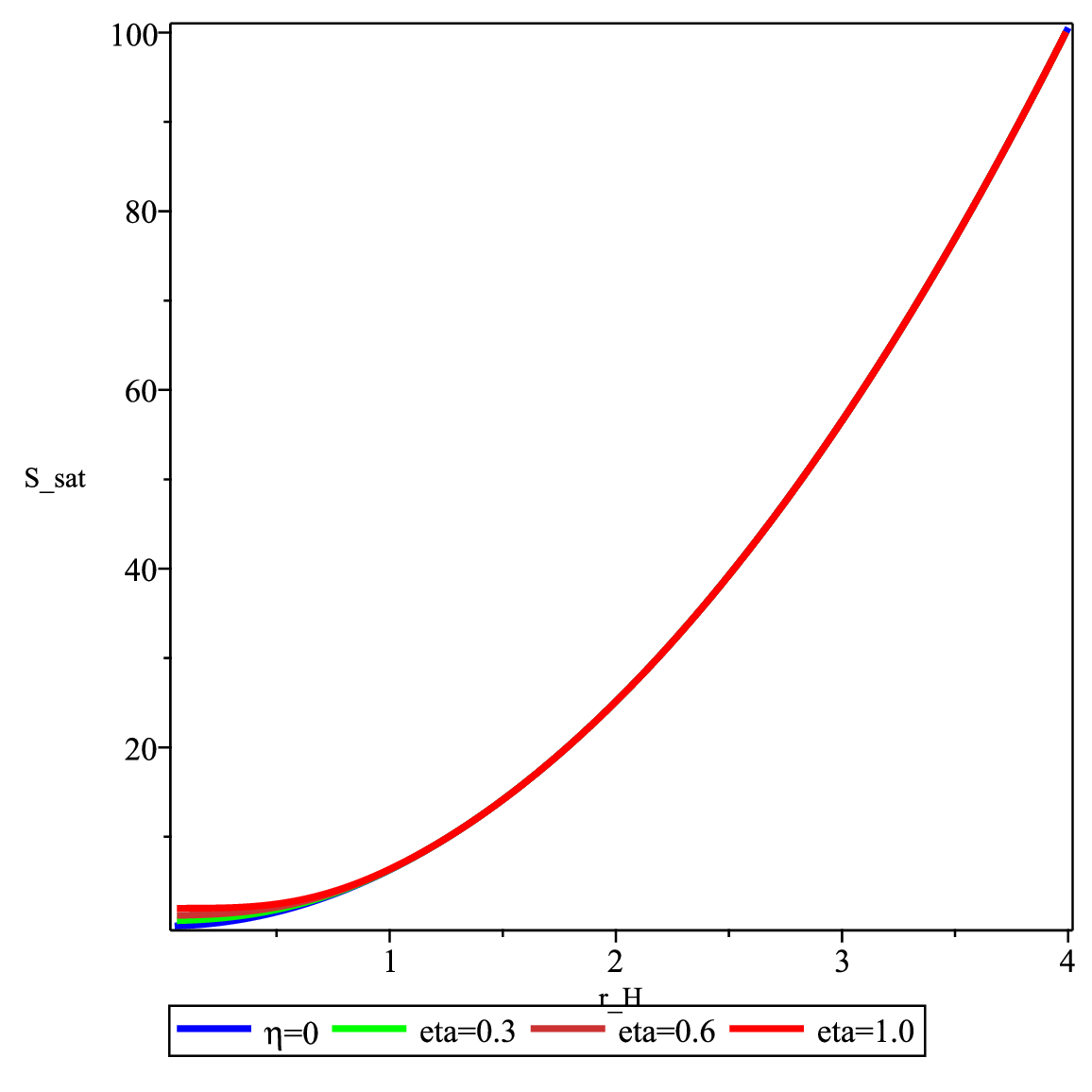}
  \caption{Saturation entropy $S_{\rm sat}$ versus $r_H$ for 
  $\eta = 0$ (blue), $0.3$ (green), $0.6$ (orange), $1.0$ (red). 
  Quantum corrections raise $S_{\rm sat}$ at small $r_H$; at large 
  $r_H$ all curves converge to $2\pi r_H^2$.}
  \label{fig20}
\end{figure}

The saturation value \eqref{eq:S_sat} sets the total entropy carried by the  radiation at late times. Figure~\ref{fig20} plots $S_{\rm sat}$ against 
$r_H$ for four values of $\eta$. The curves separate only at small $r_H$, 
where the exponential term $2\eta e^{-\pi r_H^2}$ is non-negligible; at  large $r_H$ they collapse onto $2\pi r_H^2$, as expected. A higher 
saturation value means the radiation must carry more entropy before the  black hole disappears, so quantum corrections increase the information 
content of the final Hawking radiation for microscopic black holes.

The parameter $a$ does not appear in Eq.~\eqref{eq:S_sat}, so curves for  different $a$ would coincide in Fig.~\ref{fig20}. Its imprint on 
information recovery is felt entirely through the Page time, not through  the amount of information ultimately recovered.

Taken together, Figs.~\ref{fig16}--\ref{fig20} draw a consistent 
picture: $\eta$ controls the magnitude of the quantum correction to both  the thermodynamic potentials and the saturation entropy, while $a$ governs 
the extremal structure that sets a hard upper limit on the horizon radii  at which island formation is physically possible. The two effects are 
independent in origin but intertwined in their consequences for information recovery.
\subsection{Relation Between Thermodynamic Stability and Information Recovery}
\label{subsec:thermo_information}
The results obtained in Sections~\ref{sec:thermo} and \ref{sec:page} reveal that the same non-perturbative quantum correction parameter $\eta$ simultaneously affects both the thermodynamic and information-theoretic properties of the Schwarzschild AdS black hole.

From the thermodynamic perspective, the correction modifies the entropy and consequently changes the heat capacity, free energy, and internal energy. As shown in Figs.~\ref{fig1}--\ref{fig3}, these modifications are most significant in the small-horizon-radius regime, where the exponential term $\eta e^{-\pi r_H^2}$ becomes appreciable. In particular, the correction shifts the phase-transition points, alters the extent of stable and unstable regions, and changes the preferred thermodynamic equilibrium configurations.

A remarkably similar behavior appears in the information-theoretic sector. The corrected entropy directly enters the generalized entropy through the island prescription and modifies both the Page time and the saturation entropy. As demonstrated in Figs.~\ref{fig17} and \ref{fig20}, increasing $\eta$ increases the entropy that must be accumulated before information recovery begins and consequently delays the Page transition.

An important observation is that the influence of $\eta$ is localized in the same physical regime in both sectors. For large horizon radius, the exponential correction rapidly vanishes and all thermodynamic quantities approach their classical Schwarzschild--AdS limits. Likewise, the correction to the saturation entropy becomes negligible and the information recovery process approaches the classical behavior. Therefore, the strongest deviations from classical physics occur precisely in the microscopic black-hole regime, where quantum gravitational effects are expected to dominate.

These results suggest a direct connection between thermodynamic stability and information recovery. The same quantum correction responsible for modifying phase transitions and equilibrium structures also changes the Page time and the late-time entropy of Hawking radiation. Although the thermodynamic and information-theoretic analyses are performed independently, both are governed by the corrected entropy relation~\eqref{eq:corrected_entropy}. This indicates that non-perturbative quantum effects simultaneously influence the macroscopic thermodynamic behavior and the microscopic information content of black holes.

The combined picture emerging from this work therefore supports the view that quantum corrections play a dual role: they modify the thermodynamic phase structure of the black hole while also affecting the mechanism through which information is recovered during the evaporation process. Such a connection provides additional evidence that thermodynamics and quantum information are deeply intertwined aspects of black-hole physics.

The structural connections observed between quantum-corrected thermodynamics and information recovery naturally motivate a deeper inquiry into the foundational consistency of the model. A powerful and rigorous test of consistency in modified gravity frameworks is the evaluation of universal thermodynamic relations in the extremal limit. In the next section, we extend our analysis to examine whether the universal extremality bounds remain robust when non-perturbative quantum corrections are incorporated.
\section{Universal Relation for Black Holes in Conformal Killing Gravity}\label{sec:universal}
The universal extremality condition represents a thermodynamic relation that connects entropy, temperature, and mass in the limit where a black hole becomes extremal. This relation has been verified across a broad spectrum of gravitational theories, ranging from Einstein gravity to higher-derivative extensions. Its universality stems from the fact that it does not depend on the specific details of the theory, but rather emerges from very general thermodynamic principles. In this section, we investigate whether this universal condition remains valid for black holes in Conformal Killing Gravity when a small perturbative deformation is introduced.
The static, spherically symmetric Schwarzschild-AdS solution in this theory has already been presented in the previous section, along with the corresponding mass, temperature, and entropy. We refer the reader to those expressions for the explicit forms.
To avoid confusion with the existing parameters of the model, we introduce a new independent perturbation parameter $\delta$, which represents a minimal deformation of the theory. Specifically, we take this perturbation to be a small shift in the AdS curvature radius: $l \to l + \delta$. Our objective is to compute the combination $-T (\partial S / \partial \delta)_{M}$ in the extremal limit and verify that it remains independent of $\delta$, which would serve as a signature of universality \cite{400,401,402,403,404,405,406}.
The key thermodynamic identity that we employ is
\begin{equation}
-T \left( \frac{\partial S}{\partial \delta} \right)_{M} = \left( \frac{\partial M}{\partial \delta} \right)_{S}.
\label{thermo_identity_CKG}
\end{equation}
This identity follows directly from the first law of thermodynamics, $dM = T dS$, and holds as long as the first law remains valid. Since the model contains no electric charge, the first law reduces to its simplest form. Consequently, our task reduces to computing $(\partial M/\partial \delta)_S$ in the extremal limit.
\subsection{Derivative of Entropy with Respect to the Perturbation Parameter}
Using the chain rule and the mass constraint, we obtain the derivative of the entropy with respect to $\delta$ at fixed mass. The detailed calculation is presented in Appendix A. The final result is
\begin{equation}
\left( \frac{\partial S}{\partial \delta} \right)_M = \frac{4\pi r_H^4 \left( 1 - \eta e^{-\pi r_H^2} \right)}{l^3 \left( 1 + \frac{3r_H^2}{l^2} - a r_H^4 \right)}.
\label{dS_ddelta_final_CKG}
\end{equation}
For the classical entropy, setting $\eta = 0$ gives
\begin{equation}
\left( \frac{\partial S_0}{\partial \delta} \right)_M = \frac{4\pi r_H^4}{l^3 \left( 1 + \frac{3r_H^2}{l^2} - a r_H^4 \right)}.
\label{dS0_ddelta_final_CKG}
\end{equation}
\subsection{The Universal Combination}
The universal combination is defined as the extremal limit of the product of the temperature and the entropy derivative with respect to the perturbation parameter:
\begin{equation}
\mathcal{U} = \lim_{M \to M_{\text{ext}}} \left[ -T \left( \frac{\partial S}{\partial \delta} \right)_{M} \right].
\end{equation}
The physical temperature associated with the corrected entropy is given by
\begin{equation}
T = T_0 \left( \frac{\partial S_0}{\partial S} \right) = \frac{1}{4\pi r_H} \left( 1 + \frac{3r_H^2}{l^2} - a r_H^4 \right) \frac{1}{2\pi r_H \left( 1 - \eta e^{-\pi r_H^2} \right)}.
\label{temp_corrected_CKG}
\end{equation}
Multiplying $-T$ by the derivative from Eq.~(\ref{dS_ddelta_final_CKG}) and simplifying gives
\begin{equation}
-T \left( \frac{\partial S}{\partial \delta} \right)_M = - \frac{r_H^3}{l^3}.
\label{universal_expression_CKG}
\end{equation}
The cancellation of the quantum correction parameter $\eta$ is worth emphasizing. The factor $(1 - \eta e^{-\pi r_H^2})$ appears in both the numerator and denominator and cancels exactly. The same result for the classical case is obtained by setting $\eta = 0$:
\begin{equation}
-T_0 \left( \frac{\partial S_0}{\partial \delta} \right)_M = - \frac{r_H^3}{l^3}.
\label{universal_expression_classical_CKG}
\end{equation}
\subsection{Extremality Condition and the Final Result}
The extremal limit corresponds to the vanishing of the temperature. From Eq.~(\ref{temp_corrected_CKG}), the condition $T = 0$ is satisfied when
\begin{equation}
1 + \frac{3r_{\text{ext}}^2}{l^2} - a r_{\text{ext}}^4 = 0,
\label{extremal_condition_CKG}
\end{equation}
provided that $1 - \eta e^{-\pi r_{\text{ext}}^2} \neq 0$. This condition determines the extremal horizon radius $r_{\text{ext}}$ as a function of $l$ and $a$. The quantum correction parameter $\eta$ does not enter the extremality condition, meaning the location of the extremal horizon is determined solely by the classical parameters.
Taking the extremal limit of Eq.~(\ref{universal_expression_CKG}) gives
\begin{equation}
\mathcal{U} = \lim_{r_H \to r_{\text{ext}}} \left( - \frac{r_H^3}{l^3} \right) = - \frac{r_{\text{ext}}^3}{l^3}.
\label{universal_result_CKG}
\end{equation}
\begin{table}[h]
\centering
\caption{Comparison of the universal combination for classical and corrected entropy in non-extremal and extremal regimes.}
\begin{tabular}{c c c c}
\hline
Entropy & Regime & $-T(\partial S/\partial \delta)_M$ & $\mathcal{U}$ \\
\hline
$S_0 = \pi r_H^2$ & Non-extremal & $-r_H^3/l^3$ & --- \\
$S = \pi r_H^2 + \eta e^{-\pi r_H^2}$ & Non-extremal & $-r_H^3/l^3$ & --- \\
$S_0 = \pi r_H^2$ & Extremal & --- & $-r_{\text{ext}}^3/l^3$ \\
$S = \pi r_H^2 + \eta e^{-\pi r_H^2}$ & Extremal & --- & $-r_{\text{ext}}^3/l^3$ \\
\hline
\end{tabular}
\end{table}
Several important observations emerge from this analysis. The universality of the extremality condition is preserved in Conformal Killing Gravity, despite the presence of the $r_H^4$ term in the metric function and the exponential correction in the entropy. The conformal Killing gravity parameter $a$ influences the value of $\mathcal{U}$ through its effect on the extremal radius, but does not break the universal behaviour.
The quantum correction parameter $\eta$ cancels out completely in the universal combination. This cancellation is a direct consequence of the thermodynamic identity and the first law. The exponential correction to the entropy appears in both the temperature and the entropy derivative, and these factors cancel exactly. This suggests that the universal relation is robust against non-perturbative corrections to the entropy, at least for the class of corrections considered here.
The result $\mathcal{U} = -r_{\text{ext}}^3/l^3$ follows directly from the first law of thermodynamics and does not depend on the specific details of the underlying theory. This confirms that the universal relation is a robust feature of black hole thermodynamics, valid even in modified gravity theories with quantum-corrected entropy.
The dependence on the perturbation parameter $\delta$ enters only through the combination $a l^2$. If the background parameters are such that this combination is held fixed, the universal combination becomes independent of $\delta$. This suggests that the universality of the extremality condition is preserved, but with a dependence on the AdS radius that may be absorbed into the definition of the conformal Killing gravity parameter.
The independence of $\mathcal{U}$ from $\eta$ implies that the extremality condition is stable under quantum corrections to the entropy. This stability provides a nontrivial check on the consistency of the thermodynamic description and lends support to the validity of the underlying framework. The universal relation thus constitutes a powerful consistency condition that any viable theory of gravity coupled to matter fields must satisfy.

\section{Conclusion}
\label{sec con}

In this work, we have examined the thermodynamic and information-theoretic properties of Schwarzschild--AdS black holes in conformal Killing gravity, incorporating the non-perturbative exponential entropy correction $S_{\mathrm{corr}} = \pi r_H^2 + \eta e^{-\pi r_H^2}$.

From a thermodynamic perspective, the non-perturbative correction preserves the standard area law for large black holes while significantly modifying the heat capacity, Helmholtz free energy, internal energy, and Gibbs free energy at small horizon radii. The phase-transition points shift with $\eta$, and thermal stability depends on the interplay between $\eta$, the conformal gravity parameter $a$, and the AdS radius $l$. Each parameter serves a distinct physical role: $\eta$ operates exclusively in the microscopic quantum regime; $a$ reshapes the overall phase structure, enabling van der Waals-type critical behavior for $a > 0$; and $l$ determines the thermodynamic pressure scale.

Analyzing thermodynamic deviations relative to the classical case demonstrates that the Gibbs free energy deviation $\Delta G_1 = G_S - G_0$ is negative and decays exponentially with horizon radius. Meanwhile, the Helmholtz free energy deviation $\Delta F_1$ is negative and the internal energy deviation $\Delta E_1$ is positive, both being confined to small radii. When temperature corrections are simultaneously included, $\Delta G_2$ remains negative with an increased magnitude, whereas $\Delta E_2$ vanishes identically, confirming that quantum corrections affect individual thermodynamic potentials differently. Furthermore, the Hawking--Page temperature behavior ensures that the equation of state remains independent of $\eta$, determined solely by the background geometry.

On the information-theoretic side, applying the island prescription to the corrected entropy yields a unitary Page curve. Without islands, the radiation entropy grows linearly without bound. Including an island causes the entropy to saturate at $S_{\mathrm{sat}} = 2(\pi r_H^2 + \eta e^{-\pi r_H^2})$ after the Page time, thereby resolving the information paradox. Notably, stronger quantum corrections delay information recovery: larger values of $\eta$ increase the saturation threshold and shift the Page time $t_P = 2S_{\mathrm{corr}}/(c\kappa)$ upward. The conformal gravity parameter $a$ influences information dynamics by reducing surface gravity, driving the system toward an extremal limit where evaporation halts.

A key highlight of this investigation is the universal extremality relation. By introducing a minimal perturbation to the AdS curvature radius, we computed the response combination $-T(\partial S/\partial \delta)_M$ in the extremal limit. Crucially, the quantum correction parameter $\eta$ cancels out completely, yielding the robust result $\mathcal{U} = -r_{\text{ext}}^3 / l^3$, where $r_{\text{ext}}$ is fixed by the classical extremality condition $1 + 3r_{\text{ext}}^2/l^2 - a r_{\text{ext}}^4 = 0$. This exact cancellation arises because exponential terms appear identically in both the temperature and the entropy derivative. This confirms that the universal relation is stable under non-perturbative quantum corrections, offering a strong theoretical consistency check for modified gravity frameworks.

In summary, the same underlying parameters govern phase structure, thermodynamic stability, information recovery, and extremality bounds, highlighting a unified connection between black hole thermodynamics and quantum information in conformal Killing gravity.

Several promising avenues remain for future research. Extending this analysis to charged or rotating black holes would enrich the phase space and island dynamics. Additionally, exploring holographic duals within the AdS/CFT correspondence could offer deeper insights into the boundary interpretation of island formation and modified Page curves in conformal Killing gravity.
\appendix 
\section{Detailed Derivations}
\subsection{Derivative of Mass with Respect to $l$ and $r_H$}
The mass function is given by
\begin{equation}
M = \frac{r_H}{2} \left( 1 + \frac{r_H^2}{l^2} - \frac{a}{5} r_H^4 \right).
\end{equation}
The derivative with respect to $l$ is
\begin{equation}
\frac{\partial M}{\partial l} = \frac{r_H}{2} \left( -\frac{2r_H^2}{l^3} \right) = - \frac{r_H^3}{l^3}.
\end{equation}
The derivative with respect to $r_H$ is
\begin{equation}
\frac{\partial M}{\partial r_H} = \frac{1}{2} \left( 1 + \frac{3r_H^2}{l^2} - a r_H^4 \right).
\end{equation}
\subsection{Derivative of Entropy with Respect to $r_H$}
The corrected entropy is
\begin{equation}
S = \pi r_H^2 + \eta e^{-\pi r_H^2}.
\end{equation}
The derivative with respect to $r_H$ is
\begin{equation}
\frac{\partial S}{\partial r_H} = 2\pi r_H - 2\pi \eta r_H e^{-\pi r_H^2} = 2\pi r_H \left( 1 - \eta e^{-\pi r_H^2} \right).
\end{equation}
\subsection{Derivative of $r_H$ with Respect to $\delta$ at Fixed Mass}
From the mass constraint $M(r_H, l(\delta)) = M$, we have
\begin{equation}
\left( \frac{\partial M}{\partial r_H} \right) \left( \frac{\partial r_H}{\partial \delta} \right)_M + \left( \frac{\partial M}{\partial l} \right) \left( \frac{\partial l}{\partial \delta} \right)_M = 0.
\end{equation}
Since $\partial l/\partial \delta = 1$, we obtain
\begin{equation}
\left( \frac{\partial r_H}{\partial \delta} \right)_M = - \frac{ \partial M/\partial l }{ \partial M/\partial r_H } = \frac{2 r_H^3}{l^3 \left( 1 + \frac{3r_H^2}{l^2} - a r_H^4 \right)}.
\end{equation}
\subsection{Derivative of Entropy with Respect to $\delta$ at Fixed Mass}
Using the chain rule,
\begin{equation}
\left( \frac{\partial S}{\partial \delta} \right)_M = \left( \frac{\partial S}{\partial r_H} \right) \left( \frac{\partial r_H}{\partial \delta} \right)_M,
\end{equation}
we get
\begin{equation}
\left( \frac{\partial S}{\partial \delta} \right)_M = 2\pi r_H \left( 1 - \eta e^{-\pi r_H^2} \right) \times \frac{2 r_H^3}{l^3 \left( 1 + \frac{3r_H^2}{l^2} - a r_H^4 \right)}.
\end{equation}
Simplifying,
\begin{equation}
\left( \frac{\partial S}{\partial \delta} \right)_M = \frac{4\pi r_H^4 \left( 1 - \eta e^{-\pi r_H^2} \right)}{l^3 \left( 1 + \frac{3r_H^2}{l^2} - a r_H^4 \right)}.
\end{equation}
\subsection{Extremal Limit Calculation}
The extremal limit is obtained by taking $r_H \to r_{\text{ext}}$, where $r_{\text{ext}}$ satisfies
\begin{equation}
1 + \frac{3r_{\text{ext}}^2}{l^2} - a r_{\text{ext}}^4 = 0.
\end{equation}
In this limit, the combination $-T(\partial S/\partial \delta)_M$ becomes
\begin{equation}
\mathcal{U} = \lim_{r_H \to r_{\text{ext}}} \left( - \frac{r_H^3}{l^3} \right) = - \frac{r_{\text{ext}}^3}{l^3}.
\end{equation}
The quantum correction parameter $\eta$ does not appear in this final expression.

\end{document}